\documentclass[twocolumn,a4paper,prd]{revtex4}

\usepackage{bm}
\usepackage{epsfig}
\usepackage{graphics}
\usepackage{natbib}
\usepackage{fancyh}
\usepackage[l]{floatflt}
\usepackage{epsfig}
\usepackage{amssymb}
\usepackage{latexsym}
\usepackage{times}
\usepackage{slashed}
\usepackage{upgreek}
\usepackage{amsmath}
\usepackage{amsfonts}
\usepackage{amsbsy}
\usepackage{amscd}
\usepackage{bbm}

\newcommand{\Eref}[1]{Eq.~(\ref{#1})}

\newcommand{\Sref}[1]{Sec.~\ref{#1}}
\newcommand{\Fref}[1]{Fig.~\ref{#1}}
\newcommand{\Tref}[1]{Table~\ref{#1}}
\newcommand{\cref}[1]{Ref.~\cite{#1}}

\newcommand{\hepth}[1]{{\ftn \tt hep-th/#1}}
\newcommand{\hepph}[1]{{\ftn\tt hep-ph/#1}}

\newcommand{\arxiv}[1]{{\ftn\tt  arXiv:#1}}

\newcommand{\bal}{\begin{align}}
\newcommand{\eall}{\end{align}}
\newcommand{\beqs}{\begin{subequations}}
\newcommand{\eeqs}{\end{subequations}}
\newcommand{\eec}{\end{center}}
\newcommand{\bec}{\begin{center}}

\newcommand{\eem}{\end{matrix}}
\newcommand{\bem}{\begin{matrix}}
\newcommand{\eeq}{\end{equation}}
\newcommand{\beq}{\begin{equation}}
\newcommand{\ba}{\begin{array}}
\newcommand{\ea}{\end{array}}
\newcommand{\bea}{\begin{eqnarray}}
\newcommand{\eea}{\end{eqnarray}}
\newcommand{\baq}{\begin{eqnarray}}
\newcommand{\eaq}{\end{eqnarray}}
\newcommand{\bel}{\begin{align}}
\newcommand{\eel}{\end{align}}

\newcommand\eqs[2]{Eqs.~(\ref{#1}) and (\ref{#2})}

\newcommand\eqss[3]{Eqs.~(\ref{#1}), (\ref{#2}) and (\ref{#3})}

\newcommand{\ftn}{\footnotesize}

\newcommand{\TeV}{{\mbox{\rm TeV}}}

\newcommand{\GeV}{{\mbox{\rm GeV}}}

\newcommand{\etal}{{\it et al.\/}}

\def\to{\rightarrow}

\def\llgm{\left\lgroup}
\def\rrgm{\right\rgroup}
\def\lf{\left(}
\def\rg{\right)}
\newcommand\vev[1]{\left\langle {#1} \right\rangle}
\newcommand\veva[1]{\langle {#1} \rangle}

\newcommand{\Gr}{\ensuremath{\widetilde{G}}}

\newcommand{\khi}{\ensuremath{K_{\rm H}}}
\newcommand{\kh}{\ensuremath{K_{\rm H}}}

\newcommand{\whi}{\ensuremath{W_{\rm H}}}
\newcommand{\wkhi}{\ensuremath{\wtilde K_{\rm H}}}

\newcommand{\mP}{\ensuremath{m_{\rm P}}}

\def\openone{\leavevmode\hbox{\small1\kern-3.8pt\normalsize1}}

\newcommand{\rcal}{\ensuremath{\mathcal R}}

\newcommand{\tom}{\ensuremath{\widetilde \Omega}}

\newcommand{\zm}{\ensuremath{Z_{-}}}
\newcommand{\zv}{\ensuremath{Z_{\rm v}}}
\newcommand{\tv}{\ensuremath{T_{\rm v}}}

\newcommand{\zp}{\ensuremath{Z_{+}}}

\newcommand{\nm}{\ensuremath{N_{\boldsymbol -}}}
\newcommand{\nnp}{\ensuremath{n_+}}

\newcommand{\nmp}{\ensuremath{n_\mp}}
\newcommand{\np}{\ensuremath{N_{\boldsymbol +}}}

\newcommand{\nf}{\ensuremath{N_{\rm f}}}
\newcommand{\nt}{\ensuremath{N_{\boldsymbol\pm}}}
\newcommand{\ntt}{\ensuremath{N_{\rm t}}}
\renewcommand{\ni}{\ensuremath{n_{i}}}
\newcommand{\nj}{\ensuremath{n_{j}}}

\newcommand{\what}{\ensuremath{\widehat}}
\newcommand{\wtilde}{\ensuremath{\widetilde}}

\newcommand{\zvi}{\ensuremath{Z_{{\rm v}i}}}
\newcommand{\zvj}{\ensuremath{Z_{{\rm v}j}}}
\newcommand{\zvl}{\ensuremath{Z_{{\rm v}\ell}}}
\newcommand{\zi}{\ensuremath{Z_{i}}}
\newcommand{\zj}{\ensuremath{Z_{j}}}
\newcommand{\zl}{\ensuremath{Z_{\ell}}}

\newcommand{\zml}{\ensuremath{Z_{\ell-}}}

\newcommand{\wl}{\ensuremath{w_{\ell}}}
\newcommand{\ai}{\ensuremath{a_i}}
\newcommand{\aj}{\ensuremath{a_j}}
\newcommand{\all}{\ensuremath{a_\ell}}
\newcommand{\kl}{\ensuremath{k_\ell}}
\newcommand{\ki}{\ensuremath{k_i}}
\newcommand{\kj}{\ensuremath{k_j}}

\newcommand{\vrml}{\ensuremath{{\rm
v}_\ell}}
\newcommand{\vcc}{\ensuremath{V_{\Lambda}}}

\newcommand{\vrm}{\ensuremath{{\rm v}}}
\newcommand{\vv}{\ensuremath{\mathcal{V}}}
\newcommand{\uu}{\ensuremath{\mathcal{U}}}

\newcommand{\vpm}{\ensuremath{v_{\boldsymbol\pm}}}
\newcommand{\Vpm}{\ensuremath{V_{\boldsymbol\pm}}}
\newcommand{\vip}{\ensuremath{v_{i}}}
\newcommand{\vim}{\ensuremath{v_{j}}}

\newcommand{\up}{\ensuremath{u_{\boldsymbol+}}}
\newcommand{\um}{\ensuremath{u_{\boldsymbol-}}}
\newcommand{\upm}{\ensuremath{u_{\boldsymbol\pm}}}
\newcommand{\uip}{\ensuremath{u_{i}}}
\newcommand{\uim}{\ensuremath{u_{j}}}

\newcommand{\art}{{\rm atn}}
\newcommand{\arth}{{\rm atnh}}

\newcommand{\wf}{\ensuremath{W_{\rm f}}}
\newcommand{\wnf}{\ensuremath{W_{N\rm f}}}
\newcommand{\calw}{\ensuremath{\mathcal{W}}}

\newcommand{\wpm}{\ensuremath{W_{0\boldsymbol\pm}}}
\newcommand{\ww}{\ensuremath{W_{\boldsymbol\pm}}}
\newcommand{\wwo}{\ensuremath{W_{0\boldsymbol\pm}}}
\newcommand{\wwt}{\ensuremath{W_{N\boldsymbol\pm}}}
\newcommand{\wwto}{\ensuremath{W_{0N\boldsymbol\pm}}}
\newcommand{\wip}{\ensuremath{W_{i}}}
\newcommand{\wim}{\ensuremath{W_{j}}}

\newcommand{\kppm}{\ensuremath{K_{\boldsymbol\pm}}}

\newcommand{\omp}{\ensuremath{\Omega_{\boldsymbol+}}}
\newcommand{\omm}{\ensuremath{\Omega_{\boldsymbol-}}}
\newcommand{\ompm}{\ensuremath{\Omega_{\boldsymbol\pm}}}
\newcommand{\omip}{\ensuremath{\Omega_{i}}}
\newcommand{\omim}{\ensuremath{\Omega_{j}}}
\newcommand{\omipm}{\ensuremath{\Omega_{\al\boldsymbol\pm}}}

\def\aal{{\bar\alpha}}
\def\bbet{{\bar\beta}}
\def\bz{{Z^*}}
\def\bzz{{\bar z}}
\def\al{{\alpha}}

\def\bt{{\beta}}

\def\bZ{Z^*}
\def\n{\bar{n}}

\newcommand{\mgr}{\ensuremath{m_{3/2}}}
\newcommand{\mgrt}{\ensuremath{\widetilde m_{3/2}}}
\newcommand{\mgrf}{\ensuremath{m_{3/2f}}}
\newcommand{\mgrnf}{\ensuremath{m_{3/2F}}}
\newcommand{\mgrpm}{\ensuremath{m_{3/2\boldsymbol\pm}}}
\newcommand{\mgrm}{\ensuremath{m_{3/2\boldsymbol-}}}

\newcommand{\mgru}{\ensuremath{m_{3/2{\cal U}}}}
\newcommand{\mgrfu}{\ensuremath{m_{3/2{\cal FU}}}}

\newcommand{\cp}{\ensuremath{C^{+}}}
\newcommand{\cm}{\ensuremath{C^{-}}}
\newcommand{\cfpu}{\ensuremath{C^{+}}}
\newcommand{\cfmu}{\ensuremath{C^{-}}}

\newcommand{\cfp}{\ensuremath{C_{f}^+}}
\newcommand{\cfm}{\ensuremath{C_{f}^-}}
\newcommand{\cfpm}{\ensuremath{C_{f}^\pm}}

\newcommand{\Cfp}{\ensuremath{C_{F}^+}}
\newcommand{\Cfm}{\ensuremath{C_{F}^-}}
\newcommand{\Cfpm}{\ensuremath{C_{F}^\pm}}

\newcommand{\Ctp}{\ensuremath{C_{T}^+}}
\newcommand{\Ctm}{\ensuremath{C_{T}^-}}
\newcommand{\Ctpm}{\ensuremath{C_{T}^\pm}}

\renewcommand\mtt[4]{\mbox{$\llgm\bem #1 &#2 \cr #3&
#4\eem\rrgm$}}

\renewcommand{\Cup}{\ensuremath{C_{\mathcal{U}}^+}}
\newcommand{\Cum}{\ensuremath{C_{\mathcal{U}}^-}}
\newcommand{\Cupm}{\ensuremath{C_{\mathcal{U}}^\pm}}

\newcommand{\Cfup}{\ensuremath{C_{F\mathcal{U}}^+}}
\newcommand{\Cfum}{\ensuremath{C_{F\mathcal{U}}^-}}
\newcommand{\Cfupm}{\ensuremath{C_{F\mathcal{U}}^\pm}}

\newcommand{\cupmm}{\ensuremath{C_{u\boldsymbol\pm}^-}}
\newcommand{\cupmp}{\ensuremath{C_{u\boldsymbol\pm}^+}}

\newcommand{\wo}{\ensuremath{W_0}}
\newcommand{\wopm}{\ensuremath{W_0^{\pm}}}
\newcommand{\wop}{\ensuremath{W_0^{+}}}
\newcommand{\wom}{\ensuremath{W_0^{-}}}
\newcommand{\wcc}{\ensuremath{W_{\Lambda}}}
\newcommand{\gk}{\ensuremath{g_{K}}}

\newcommand{\Kpm}{\ensuremath{K_{\boldsymbol\pm}}}
\newcommand{\Kp}{\ensuremath{K_{\boldsymbol+}}}
\newcommand{\Km}{\ensuremath{K_{\boldsymbol-}}}

\newcommand{\phc}{\ensuremath{\Phi}}

\def\Ka{K\"{a}hler potential}
\def\Kmn{K\"{a}hler manifold}
\def\Kaa{K\"{a}hler~}

\newcommand{\plk}{{\it Planck}}

\newcommand{\Tr}{\mbox{\sf Tr}}
\newcommand{\diag}{\ensuremath{{\sf diag}}}

\newcommand{\re}{\ensuremath{{\sf Re}}}

\renewcommand{\refname}{{\bf\scshape References}}

\renewcommand{\thesubsection}{{\small\sf\Alph{subsection}}}

\renewenvironment{subequations}{%
\refstepcounter{equation}%
\setcounter{parentequation}{\value{equation}}%
  \setcounter{equation}{0}
  \def\theequation{\theparentequation{\sffamily\ftn\alph{equation}}}%
  \ignorespaces
}{%
  \setcounter{equation}{\value{parentequation}}%
  \ignorespacesafterend
}

  \makeatletter
\renewcommand*\l@section[2]
  {%
    \ifnum \c@tocdepth >\z@
      \addpenalty \@secpenalty
      \addvspace {1.em \@plus \p@ }%
      \setlength \@tempdima {1.8em}%
      \begingroup
        \parindent \z@
        \rightskip \@pnumwidth
        \parfillskip -\@pnumwidth
        \leavevmode \bfseries
        \advance \leftskip \@tempdima
        \hskip -\leftskip
        #1\nobreak
        \hfil
        \nobreak
        \hb@xt@ \@pnumwidth {\hss #2\kern -\p@ \kern \p@ }%
        \par
      \endgroup
    \fi
  }
\makeatother

\begin{document}

\title{\bf\scshape  From Minkowski to de Sitter Vacua with Various Geometries}

\author{{\scshape Constantinos Pallis} \\ {\small\it Laboratory of Physics, Faculty of
Engineering, Aristotle University of Thessaloniki, GR-541 24
Thessaloniki, GREECE} \\ {\ftn\sl  e-mail address: }{\ftn\tt
kpallis@gen.auth.gr}}

\begin{abstract}

\noindent {\ftn \bf\scshape Abstract:} New no-scale supergravity
models with F-term SUSY breaking are introduced, adopting \Ka s
parameterizing flat or curved (compact or non-compact) \Kmn s. We
systematically derive the form of the superpotentials leading to
Minkowski vacua. Combining two types of these superpotentials we
can also determine de Sitter or anti-de Sitter vacua. The
construction can be easily extended to multi-modular settings of
mixed geometry. The corresponding soft SUSY-breaking parameters
are also derived.
\\ \\ {\scriptsize {\sf PACs numbers: 12.60.Jv, 04.65.+e}
%11.30.Er, 11.30.Pb,

%\hfill {\sl\bfseries Published in} {\sl Phys. Rev. D} {\bf 86},
%023523 (2012)

}
%\pacs{98.80.Cq, 11.30.Qc, 12.60.Jv} it does not require fine tuned parameters

\end{abstract}\pagestyle{fancyplain}

\maketitle

\rhead[\fancyplain{}{ \bf \thepage}]{\fancyplain{}{\sl From
Minkowski to de Sitter Vacua with Various Geometries}}
\lhead[\fancyplain{}{\sl C. Pallis}]{\fancyplain{}{\bf \thepage}}
\cfoot{}

\noindent\rule\columnwidth{.4pt}\vspace*{-.5cm}{\tableofcontents}\noindent\rule\columnwidth{.4pt}\\
\section{\bfseries\scshape Introduction}\label{intro}

Within \emph{Supergravity} ({\ftn\sf SUGRA}) \cite{nilles, mass},
breaking \emph{Supersymmetry} ({\ftn\sf SUSY}) on a sufficiently
flat background requires a huge amount of fine tuning, already at
the classical level -- see e.g. \cref{polonyi, hall}. Besides
remarkable exceptions presented recently \cite{susyr}, the
so-called no-scale models \cite{noscale, old, noscale18,
noscale19, Rgauged,nsreview, fabio, burgess, roest} provide an
elegant framework which alleviates the problem above since SUSY is
broken with naturally vanishing vacuum energy along a flat
direction.  On the other hand, the discovery of the accelerate
expansion of the present universe \cite{plcp} motivates us to
develop models with \emph{de Sitter} ({\ftn\sf dS}) -- or even
\emph{anti-dS} ({\ftn\sf AdS}) -- vacua which may explain this
expansion -- independently of the controversy \cite{vafa, lindev,
sevilla, ferara} surrounding this kind of (meta) stable vacua
within string theory.

In two recent papers \cite{noscale18, noscale19}, a systematic
derivation of dS/AdS vacua is presented in the context of the
no-scale SUGRA without invoking any external mechanism of vacuum
uplifting such as through the addition of anti-D3 brane
contributions \cite{kallosh} or extra Fayet-Iliopoulos terms
\cite{antst}. Namely, these vacua are achieved by combining two
distinct Minkowski vacua taking as initial point the \Ka\
parameterizing the non-compact $SU(1,1)/U(1)$ \Kmn\ in half-plane
coordinates, $T$ and $T^*$. Possible instabilities along the
imaginary direction of the $T$ field can be cured by introducing
mild deformations of the adopted geometry. The analysis has been
extended to incorporate more than one superfields in conjunction
with the implementation of observationally successful inflation
\cite{noscaleinfl,nsreview}.

In this paper we show that the method above has a much wider
applicability since it remains operational for flat spaces or
curved ones. This is possible since the no-scale ``character'' of
the models, as defined above, stems from the existence of a flat
direction with SUSY broken along it, and not from the adopted
moduli geometry.  We parameterize the curved spaces of our models
with the Poincar\'e disk coordinates $Z_\al$ and $\bZ_\aal$ which,
although are widely adopted within the inflationary model building
\cite{linde, alinde, sor}, they are not frequently employed for
establishing SUSY-breaking models -- cf.~\cref{susyr, Rgauged}.
This  parametrization gives us the opportunity to go beyond the
non-compact geometry \cite{noscale18, noscale19} and establish
SUSY-breaking scenaria with compact \cite{su11} or ``mixed''
geometry. In total, we here establish three novel uni-modular
no-scale models and discuss their extensions to the multi-modular
level. In all cases, we show that a subdominant quartic term
\cite{noscale18, susyr} in the \Ka\ stabilizes the sgoldstino
field to a specific vacuum and provides mass for its scalar
component without disturbing, though, the constant vacuum energy
density. This can be identified with the present cosmological
constant by finely tuning one parameter of the model whereas the
others can be adjusted to perfectly natural values. If we connect,
finally, our hidden sectors with some sample observable ones,
non-vanishing \emph{soft SUSY-breaking} ({\sf\ftn SSB}) parameters
\cite{soft}, of the order of the gravitino mass can be readily
determined at the tree level.

We start our presentation with a simplified generic argument which
outlines the transition from Minkowski to dS/AdS vacua in
\Sref{sup}. We then detail our models adopting first -- in
\Sref{fl} -- flat moduli geometry and then -- see \Sref{cu} -- two
versions of curved geometry. Generalization of our findings
displaying multi-modular models with mixed geometry is presented
in \Sref{mx}. We also study in \Sref{obs} the communication of the
SUSY breaking to the observable sector by computing the SSB terms.
We summarize our results in \Sref{con}. Some useful formulae
related to derivation of mass spectra in SUGRA with dS/AdS vacua
is arranged in Appendix~\ref{app1}. In Appendix \ref{hp} we show
the consistency of our results with those in \cref{noscale18,
noscale19} translating the first ones in the language of the
$T-T^*$ coordinates.

Unless otherwise stated, we use units where the reduced Planck
scale $\mP=2.4\cdot 10^{18}~\GeV$ is taken to be unity and the
star ($^*$) denotes throughout complex conjugation. Also, no
summation convention is applied over the repeated Latin indices
$\ell, i$ and $j$.

% -- cf. \cref{dvali, univ}To some extent,
%Although quite compelling, the above scheme gets into trouble due
%to the presence of the massless $R$ axion.

\section{\bfseries\scshape Start-Up Considerations} \label{sup}

The generation mechanism of dS/AdS vacua from a pair of Minkowski
ones can be roughly established, if we consider a uni-modular
model without specific geometry. In particular, we adopt a \Ka\
$K=K(Z,Z^*)$ and attempt to determine an expression for the
superpotential $W=W(Z)$ so as to construct a no-scale scenario.

The SUGRA potential $V$ based on $K$ and $W$ from \Eref{Vsugra} is
written as
\beq V=e^K\lf \gk^{-1}\left |\partial_Z W+W
K_Z\right|^2-3|W|^2\rg,\label{Vgen}\eeq
where $\gk^{-1}=K_{ZZ^*}^{-1}=K^{ZZ^*}$. Suppose that there is an
expression $W=\wo(Z)$ which assures that the direction $Z=Z^*$ is
classically flat with $V=0$, i.e., it provides a continuum of
Minkowski vacua. The determination of $\wo$, based on \Eref{Vgen},
entails
\beq \gk^{-1}\lf \wo'+\wo
K_Z\rg^2=3\wo^2~\Rightarrow~\frac{d\wo}{dZ\wo}=\pm\sqrt{3\gk}-K_Z\label{wode}\eeq
with \Eref{ext} being satisfied -- the relevant conditions may
constrain the model parameters once $K$ is specified. Here prime
stands for derivation \emph{with respect to} ({\ftn\sf w.r.t})
$Z$. \Eref{wode} admits obligatorily two solutions
\beq \wopm=m\exp\lf \pm\int dZ\sqrt{3\gk}-\int
dZK_Z\rg\label{wosol}\eeq
with $m$ a mass parameter. For the $K$'s considered below, it is
easy to verify that
\beq \int dZ K_Z = K/2 \label{khf}\eeq
up to a constant of integration. E.g., if $K=|Z|^2$, then $K_Z=Z$
for $Z^*=Z$ and $\int dZ K_Z=Z^2/2=K/2$.

According to \cref{noscale18, noscale19}, the appearance of dS/AdS
vacua is attained, if we consider the following linear combination
of $\wopm$ in \Eref{wosol}
\beq \wcc=\cp\wop-\cm\wom, \label{wcc}\eeq
where $\cm$ and $\cp$ are non-zero constants. As can be easily
checked, $\wcc$ does not consist solution of \Eref{wode}. It
offers, however, the achievement of a technically natural dS/AdS
vacuum since its substitution into \Eref{Vsugra} yields
\beq \begin{aligned} \vcc&= e^K\lf \gk^{-1}\lf\wcc'+\wcc K_Z\rg^2
-3\wcc^2\rg\\ &=12e^K\cm\cp\wom\wop=12m^2\cm\cp,
\end{aligned}\label{Vlf}\eeq
where we take into account \eqs{wosol}{khf}. Rigorous validation
and extension (to more superfields) of this method can be
accomplished via its application to specific working models. This
is done in the following sections.

Let us, finally, note that $\vcc$ can be identified with the
present cosmological constant by demanding
\beq \label{omde} \vcc=\Omega_\Lambda\rho_{\rm
c0}=7.2\cdot10^{-121}\mP^4,\eeq
where $\Omega_\Lambda=0.6889$ and $\rho_{\rm
c0}=2.4\cdot10^{-120}h^2\mP^4$ with $h=0.6732$ \cite{plcp} is the
density parameter of dark energy and the current critical energy
density of the universe.

\section{\bfseries\scshape Flat Moduli Geometry} \label{fl}

We focus first on the models with flat internal geometry and
describe below their version for one -- see \Sref{fl1} -- or more
-- see \Sref{fl2} -- moduli.

\subsection{\sc\small\sffamily  Uni-Modular Model} \label{fl1}

Our initial point is the \Ka\
\beq K_{\rm f}=|Z|^2-k^2\zv^4\label{Kf}\eeq
where we include the stabilization term
\beq \label{Zv} \zv=\zp-\sqrt{2}{\rm
v}~~\mbox{with}~~Z_\pm=Z\pm\bZ. \eeq
Here $k$ and $\vrm$ are two real free parameters. Small $k$ values
are completely natural, according to the 't Hooft argument
\cite{symm}, since $K_{\rm f}$ enjoys an enhanced $U(1)$ symmetry
which is exact for $k=0$. It is evident that the $Z$ space defined
by $K_{\rm f}$ is flat with metric $\vev{K_{Z\bZ}}=1$ along the
stable configurations
\beqs\bea \label{vev1} & Z=\bZ&~~\mbox{for}~~k=0~~\\
\mbox{and}~~&\vev{Z}=\vrm/\sqrt{2}&~~\mbox{for}~~k\neq0.\label{vev2}
\eea\eeqs
Hereafter, the value of a quantity $Q$ for both alternatives
above, -- i.e. either along the flat direction of \Eref{vev1} or
at the (stable) minimum of \Eref{vev2} -- is denoted by the same
symbol $\vev{Q}$.

Applying \Eref{Vsugra} for $K=K_{\rm f}$, and an unknown $W=\wf$
for $Z=\bz$, we obtain
\beq V_{\rm f}=e^{Z^2}\lf\lf
Z\wf+\wf'\rg^2-3\wf^2\rg.\label{Vf0}\eeq
Following the strategy in \Sref{sup}, we first find the required
form of $W_{\rm f}$,  $W_{0\rm f}$, which assures the
establishment of a $Z$-flat direction with Minkowski vacua. I.e.
we require $\vev{V_{\rm f}}=0$ for any $Z$. Solving the resulting
ordinary differential equation
\beq  Z+\frac{d W_{0{\rm f}}}{dZ W_{0\rm
f}}=\pm\sqrt{3}\label{Wde}\eeq
w.r.t $W_{0\rm f}$, we obtain two possible forms of $W_{0\rm f}$,
\beq W_{0\rm
f}^{\pm}(Z)=mwf^{\pm1}~~\mbox{with}~~w=e^{-Z^2/2}~~\mbox{and}~~f=e^{\sqrt{3}Z}.
\label{W0}\eeq
Note that a factor $w$ appears already in the models of
\cref{noscale19} associated, though, with a matter field and not
with the goldstino superfield as in our case.

The solutions in \Eref{W0} above can be combined as follows -- cf.
\Eref{wcc} --
\beq W_{\Lambda\rm f}=\cp W_{0\rm f}^{+}-\cm W_{0\rm f}^{-}=
mwf\cfm, \label{Wl}\eeq
where we introduce the symbols
\beq \cfpm:=\cp\pm\cm f^{-2}.\label{cfpm}\eeq
Employing $K=K_{\rm f}$ and $W=W_{\Lambda\rm f}$ from
\eqs{Kf}{Wl}, we find the corresponding $V$ via \Eref{Vsugra}
\beq \begin{aligned} V_{\rm f}&=\lf m^2/(1-12k^2\zv^2)\rg|f|^2\exp\lf-\zm^2/2-k^2\zv^4\rg\\
&\cdot\lf\left|\sqrt{3}\cfp-\zm\cfm-4k^2\zv^3\cfm\right|^2-3\left|\cfm\right|^2\rg,\end{aligned}
\label{Vf1}\eeq
which exhibits the dS/AdS vacua in \eqs{vev1}{vev2}. Indeed, we
verify that $\vev{V_{\rm f}}= \vcc$, given in \Eref{Vlf} and
\Eref{ext} for $V=V_{\rm f}$ and $\al=1$ is readily fulfilled.
Indeed, decomposing $Z_\al$ (with $\al=1$ suppressed when we have
just one $Z$) in real and imaginary parts, -- $z_1:=z$ and
$\bzz_1:=\bzz$ -- i.e.,
\beq Z_\al=(z_\al+i\bzz_\al)/\sqrt{2},\label{Zzzi}\eeq
we find that the eigenvalues of $M_0^2$ in \Eref{mbos} are
\beq \what m^2_{z\rm
f}=144k^2\mgrf^2\vev{{\cfp}/{\cfm}}^2~~\mbox{and}~~\what
m_{\bzz\rm f}=4\mgrf^2,\label{mzz}\eeq
where $\mgrf$ is the \Gr\ mass along the configurations in
\eqs{vev1}{vev2}. This is found by replacing $K$ and $W$ from
\eqs{Kf}{Wl} in \Eref{mgr}, with result
\beq \label{mgrf} \begin{aligned}\mgrf&=m\vev{f\cfm}\\
&=m\begin{cases} e^{\sqrt{3}Z}\lf\cfpu-\cfmu e^{-2\sqrt{3}Z}\rg
&\mbox{for}~~k=0\\  e^{\sqrt{3/2}\vrm}\lf\cfpu-\cfmu
e^{-\sqrt{6}\vrm}\rg&\mbox{for}~~k\neq0.\end{cases}\end{aligned}\eeq
Note that, for $k=0$ and unfixed $Z$, $\mgrf$ remains undetermined
validating thereby the no-scale character of our models -- cf.
\cref{old, nsreview}. As a shown in \Eref{mzz}, the real component
$z$ of $Z$ remains massless due to the flatness of $V_{\rm f}$
along the direction in \Eref{vev1}. However, the $k$-dependent
term in \Eref{Kf} not only stabilizes $Z$ but also provides mass
to $z$. On the other hand, this term generates poles and so
discontinuities in $V_{\rm f}$ -- see \Eref{Vlf}. We are obliged,
therefore, to focus on a local dS/AdS minimum as in \Eref{vev2}.
Inserting \eqs{mzz}{mgrf} into \Eref{strace} we find
\beq {\sf STr}M_{\rm f}^2 =\what m^2_{z\rm f}+\what m^2_{\bzz\rm
f}-4\mgrf^2=\what m^2_{z\rm f},\label{tr0}\eeq
which is consistent with \Eref{strace1} given that ${\cal R}$ in
\Eref{rcal} is found to be $\vev{\rcal_{\rm f}}=24k^2$.

%%%%%%%%%%%%%%%%%%%%%%%%%%%%%%%%%%%%%%%%%%%%%%%%%%%%%%%%%%%%%%%%%%%%
\begin{figure}[t]%
\epsfig{file=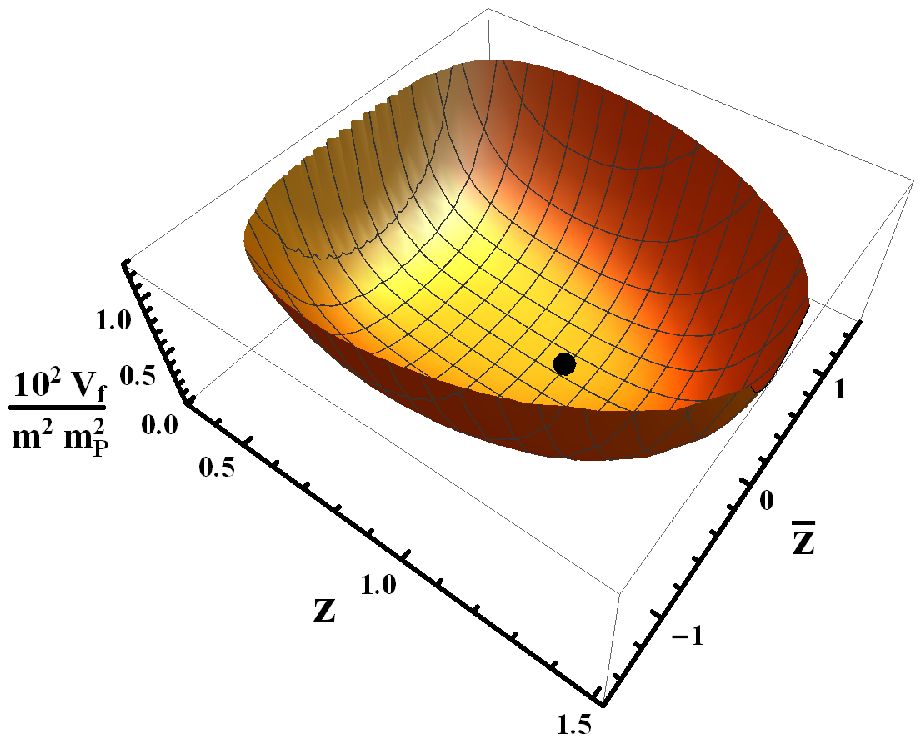,width=8.7cm,,angle=-0}
\vspace*{2.3in} \caption{\sl \small The (dimensionless) SUGRA
potential $10^2V_{\rm f}/m^2\mP^2$ in \Eref{Vf1} as a function of
$z$ and $\bzz$ in \Eref{Zzzi} for the inputs shown in column A of
\Tref{tab}. The location of the dS vacuum in \Eref{vev2} is also
depicted by a thick black point.}\label{fig1}
\end{figure}
%%%%%%%%%%%%%%%%%%%%%%%%%%%%%%%%%%%%%%%%

Our analytic findings above can be further confirmed by
\Fref{fig1}, where the dimensionless quantity $10^2V_{\rm
f}/m^2\mP^2$ is plotted as a function of $z$ and $\bzz$ in
\Eref{Zzzi}. We employ the values of the parameters listed in
column A of \Tref{tab} -- obviously $k_1$ there is identified with
$k$ in \Eref{Kf}. We see that the dS vacuum in \Eref{vev2} --
indicated by the black thick point -- is placed at
$(z,\bzz)=(1,0)$ and is stabilized w.r.t both directions. In the
same column of \Tref{tab} displayed are also the various masses of
$\Gr$ and the scalar ($\what z$) and pseudoscalar
($\what{\bar{z}}$) components of the sgoldstino $Z$ given in \GeV\
for convenience. For $N=1$, the spectrum does not comprise any
goldstino $(\what{\tilde{z}})$ as explained in
Appendix~\ref{app1}. It is worth mentioning that the
aforementioned masses may acquire quite natural values (of the
order of $10^{-15}$) for logical values of the relevant parameters
despite the fact that the fulfilment of \Eref{omde} via \Eref{Vlf}
requires a tiny $\cm$. E.g., for the parameters given in
\Tref{tab} we need $\cm=1.4\cdot 10^{-90}$.

Let us, finally, note that performing a \Kaa transformation
\beq \label{Kst} K\to
K+\Lambda_K+\Lambda_K^*\>\>\>\mbox{and}\>\>\>W\to
We^{-\Lambda_K},\eeq
with $K=K_{\rm f}$ and $W=W_{\Lambda\rm f}$ in \eqs{Kf}{Wl}
respectively and $\Lambda_K=-Z^2/2$, the present model is
equivalent with that described by the following $K$ and $W$
\beq\label{sym} \widetilde K_{\rm
f}=-\frac12\zm^2-k^2\zv^4~~\mbox{and}~~\widetilde W_{\Lambda\rm
f}= mf\cfm. \eeq
From the form above we can easily infer that, for $k\to0$,
$\widetilde K_{\rm f}$ enjoys the enhanced symmetries
\beq\label{sym1} Z \to Z + c ~~\mbox{and}~~ Z \to -Z, \eeq
where $c$ is a real number. These are more structured than the
simple $U(1)$ mentioned below \Eref{Kf} and underline, once more,
the naturality of the possible small $k$ values. In this limit, a
similar model arises in the context of the $\al$-scale SUGRA
introduced in \cref{roest}.

%%%%%%%%%%%%%%%%%%%%%%%%%%%%%%%%%%%%%%%%%%%%%%%%%%%%%%%%%%%%%%%%%%%
\begin{table}[t!]
\caption{\sl A Case Study Overview }
\begin{ruledtabular}
\begin{tabular}{c||c|c|c|c|c|c}
%\toprule
{\sc Case:}&A&B&C&D&E&F\\ \hline\multicolumn{7}{c}{\sc Input
Settings}
\\ \hline
$K$ &$K_{\rm f}$&$K_{2\rm
f}$&$K_{\boldsymbol+}$&$K_{\boldsymbol-}$&$K_{\boldsymbol+-}$&$K_{\rm
f\boldsymbol-}$\\
$W$&$W_{\Lambda\rm f}$&$W_{\Lambda2\rm
f}$&$W_{\Lambda\boldsymbol+}$&$W_{\Lambda\boldsymbol-}$&$W_{\Lambda\boldsymbol+-}$
&$W_{\Lambda\rm f\boldsymbol-}$\\
\hline \multicolumn{7}{c}{\sc Input Parameters}\\ \hline
\multicolumn{7}{c}{$\vrm_\al=\mP$, $\cp=0.01$ and $m=5~\TeV$} \\
\hline
$k_1$&$0.3$&{$0.3$}&$0.3$&$0.3$&{$0.3$}&$0.3$\\
$k_2$&$-$&$0.2$&$-$&$-$&$0.2$&$0.2$\\
$a_1$&$-$&$1$&$-$&$-$&$1$&$1$\\
$n_1$&$-$&$-$&$1$&$4$&$1$&$4$\\
$n_2$&$-$&$-$&$-$&$-$&$4$&$-$\\\hline
\multicolumn{7}{c}{\sc Particle Masses in $\GeV$} \\ \hline
$\Gr$&$170$&$276$&$145$&$180$&$263$&$288$\\\hline
$\what{z}_1$&$612$&$572$&$960$&$530$&$744$&$599$\\
$\what{\bzz}_1$&$340$&{$551$}&$581$&$180$&$1005$&$577$\\
$\what{z}_2$&$-$&$540$&$-$&$-$&$422$&$528$\\
$\what{\bzz}_2$&$-$&$551$&$-$&$-$&$372$&$466$\\\hline
$\what{\tilde z}_1$&$-$&$276$&$-$&$-$&$263$&$288$\\%\botrule
\end{tabular}\label{tab}
\end{ruledtabular}
\end{table}

%%%%%%%%%%%%%%%%%%%%%%%%%%%%%%%%%%%%%%%%%%%%%%%%%%%%%%%%%%%%%%5

\subsection{\sc\small\sffamily  Multi-Modular Model} \label{fl2}

The model above can be extended to incorporate more than one
modulus. In this case, the corresponding $K$ is written as
\beq K_{N\rm
f}=\sum_{\ell=1}^{\nf}\lf|Z_\ell|^2-\kl^2\zvl^4\rg\label{Kfn}\eeq
where for any modulus $Z_{\al}$ we include a stabilization term
\beq \label{Zvi} Z_{\rm v\al}=Z_{\al+}-\sqrt{2}{\rm
v}_\al~~\mbox{with}~~Z_{\al\pm}=Z_\al\pm\bZ_\al, \eeq
with $\al=\ell$ in the domain of the values shown in \Eref{Kfn}.
As we verify below, for the same $\al$ values, we can obtain the
stable configurations
\beqs\bea &Z_{\al-}=0 &~~\mbox{for}~~k_\al=0 \label{vevi1}\\
\mbox{and} &\vev{Z_\al}={\rm v}_\al/\sqrt{2}&
~~\mbox{for}~~k_\al\neq0. \label{vevi2}\eea\eeqs
Along them the K\"ahler metric is represented by a $\nf\times\nf$
diagonal matrix
\beq \vev{K_{\al\bbet}}=\diag(1,...,1).\eeq

Setting $\zl=Z_\ell^*$, $K=K_{N\rm f}$ from \Eref{Kfn} and
$W=\wnf(\zl)$ in \Eref{Vsugra}, $V$ takes the form
\beq V_{N\rm f}=e^{\sum_\ell Z_\ell^2}\lf\mbox{$\sum_\ell$}\lf
Z_\ell\wnf+\partial_\ell\wnf\rg^2-3\wnf^2\rg.\label{Vf0n}\eeq
Setting $V_{N\rm f}=0$ and assuming the following form for the
corresponding $\wnf$
\bea\nonumber W_{0N\rm f}(Z_1,...,Z_{\nf})&=&\prod_{\ell} W_{{\rm
f}\ell}(\zl)\Rightarrow\\ \partial_\ell W_{0N\rm f}&=&{dW_{{\rm
f}\ell}\over d\zl}{W_{0N\rm f}\over W_{{\rm f}\ell}},
\label{Wft}\eea
we obtain the separated differential equations
\beq \sum_\ell\lf \zl+\frac{d W_{{\rm f}\ell}}{d\zl W_{{\rm
f}\ell}}\rg^2=3.\label{Wdei}\eeq
These can be solved w.r.t $W_{{\rm f}\ell}$, if we set
\beq \zl+\frac{d W_{{\rm f}\ell}}{d\zl W_{{\rm
f}\ell}}=|\all|~~\mbox{with}~~ \sum_\ell\all^2=3,\label{aif} \eeq
i.e., the $\all$'s satisfy the equation of the hypersphere
$\mathbb{S}^{\nf-1}$ with radius $\sqrt{3}$. The resulting
solutions take the form
\beq W_{{\rm f}\ell}^{\pm}=\wl
f_\ell^{\pm{\all}/{\sqrt{3}}}~~\mbox{with}~~\wl=e^{-\zl^2/2}~~\mbox{and}~~f_\ell=e^{\sqrt{3}\zl}.
\label{W0i}\eeq
The total expression for $W_{0N\rm f}$ is found substituting the
findings above into \Eref{Wft}. Namely,
\beq W_{0N\rm f}^{\pm}=m\prod_{\ell} W_{{\rm f}\ell}^{\pm}=m{\cal
W}F^{\pm1},  \label{Wnf0}\eeq
where we define the functions
\beq {\cal W}=e^{-\sum_\ell
\zl^2/2}~~\mbox{and}~~F=e^{\sum_\ell\all\zl}. \label{wfi}\eeq

As in the case with $\nf=1$, we combine both solutions above as
follows
\bea W_{\Lambda N\rm f}=\cp W_{0N\rm f}^{+}-\cm W_{0N\rm f}^{-}=
m{\cal W}F\Cfm,\label{Wll}\eea
where we introduce the ``generalized'' $C$ symbols -- cf.
\Eref{cfpm}
\beq \Cfpm:=\cp\pm\cm F^{-2}.\label{cfpmi}\eeq
Substituting \eqs{Kfn}{Wll} into \Eref{Vsugra} we find that $V$
takes the form
\beq\begin{aligned} V_{N\rm f}&=m^2|F|^{2}\exp\lf-\sum_\ell\lf\zml^2/2+\kl^2\zvl^4\rg\rg\\
&\cdot\lf\sum_\ell\frac{\left|\all\Cfp-\zml\Cfm-4\kl^2\zvl^3\Cfm\right|^2}{1-12\kl^2\zvl^2}-3\left|\Cfm\right|^2\rg.
\label{Vnf}\end{aligned}\eeq
We can confirm that $V_{N\rm f}$ admits the dS/AdS vacua in
\eqs{vevi1}{vevi2} for $\al=\ell$, since $\vev{V_{N\rm f}}=\vcc$,
given in \Eref{Vlf}. In addition, \Eref{ext} for $V=V_{N\rm f}$
and $\al=\ell$ is satisfied, since the $2\nf$ masses squared of
the relevant matrix in \Eref{mbos} are positive. Indeed, analyzing
$Z_\ell$ in real and imaginary parts as in \Eref{Zzzi}, we find
\beqs\bea\what m_{z\rm f\ell}^2&=&48\kl^2
a_\ell^2\mgrnf^2\vev{{\Cfp}/{\Cfm}}^2; \label{mzi}\\ \what m_{\bar
z\rm f
\ell}^2&=&4\mgrnf^2\lf1+(3-\all^2)\vcc/6{\mgrnf^2}\rg,\label{mbzi}\eea\eeqs
where $\mgr$ for the present case is computed inserting
\eqs{Kfn}{Wll} into \Eref{mgr} with result
\beq \label{mgrfN} \begin{aligned}&\hspace*{2cm}\mgrnf=m\vev{F\Cfm}=\\
&m\cdot\begin{cases} e^{\sum_\ell\all\zl}\lf\cfpu-\cfmu
e^{-2\sum_\ell\all\zl}\rg &\mbox{for}~~\kl=0\\
e^{\sum_\ell\all\frac{\vrml}{\sqrt{2}}}\lf\cfpu-\cfmu
e^{-\sqrt{2}\sum_\ell\all\vrml}\rg&\mbox{for}~~\kl\neq
0.\end{cases}\end{aligned}\eeq
The expressions above conserve the basic features of the no-scale
models as explained below \Eref{mgrf}. We consider the stabilized
version of these models (with $k_\al\neq0$) as more complete since
it offers the determination of $\mgr$ and avoids the presence of a
massless mode which may be problematic.

We should note that the relevant $2\nf\times2\nf$ matrix $M_0^2$
of \Eref{mbos} turns out to be diagonal up to some tiny mixings
appearing in the $\bzz_\ell-\bzz_{\bar\ell}$ positions.  These
contributions though can be safely neglected since these are
proportional to $\vcc$. We also obtain $\nf-1$ Weyl fermions with
masses $\what m_{\tilde z\tilde \ell}=\mgr$ where
$\tilde\ell=1,...,\nf-1$. Note that \eqss{mgrfN}{mzi}{mbzi} reduce
to the ones obtained for $\nf=1$, i.e. \eqs{mgrf}{mzz}, if we
replace $\all=\sqrt{3}$. Inserting the mass spectrum above into
\Eref{strace}, we find
\beq{\sf STr}M_{N\rm
f}^2=2(\nf-1)\lf\mgrnf^2+V_{\Lambda}\rg+\mbox{$\sum_\ell$}\what
m^2_{z\rm f \ell}.\label{trN}\eeq
%2
This result is consistent with \Eref{strace2} given that its last
term turns out to be equal to the last term of \Eref{trN}.

\begin{figure}[t]%
\epsfig{file=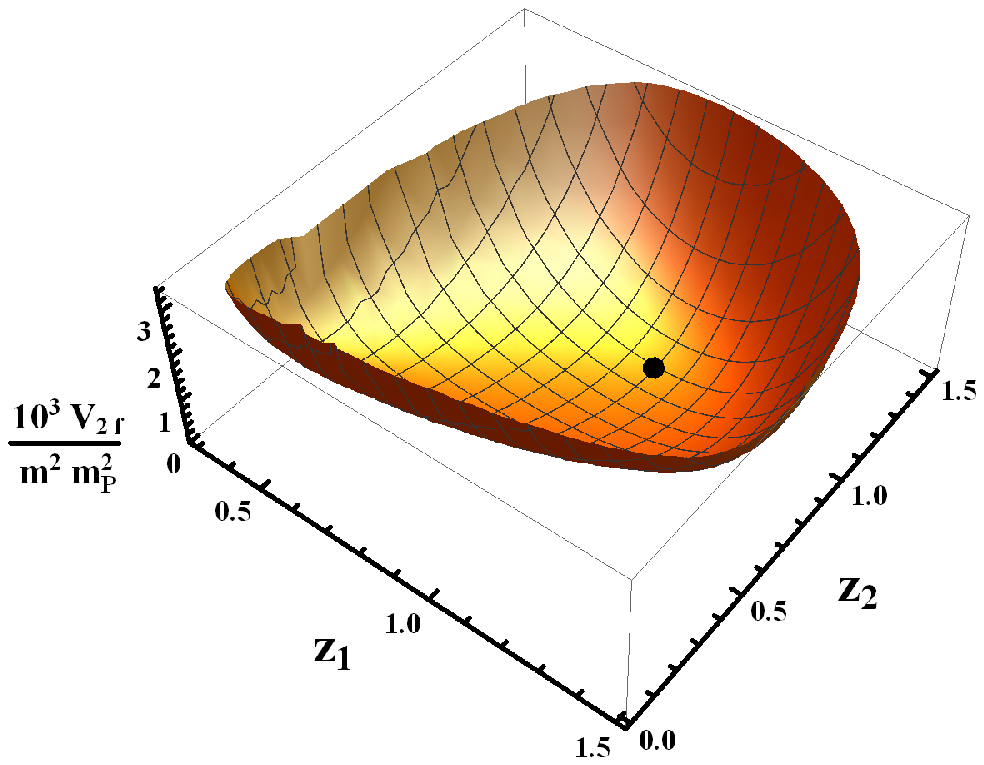,width=8.7cm,,angle=-0}
\vspace*{2.3in} \caption{\sl \small The (dimensionless) SUGRA
potential $V_{2\rm f}/m^2 \mP^2$ in \Eref{Vnf} as a function of
$z_1$ and $z_2$ in \Eref{Zzzi} for $\bzz_1=\bzz_2=0$ and the
inputs shown in column B of \Tref{tab}. The location of the dS
vacuum in \Eref{vevi2} is also depicted by a thick black
point.}\label{fig2}
\end{figure}

To highlight further the conclusions above we depict in
\Fref{fig2} the dimensionless $V_{N\rm f}$ for $N=2$, i.e.
$V_{2\rm f}$, as a function of $z_1$ and $z_2$ for
$\bzz_1=\bzz_2=0$ and the other parameters displayed in column B
of \Tref{tab}. We observe that the dS vacuum in \Eref{vevi2} --
indicated by a black thick point in the plot -- is well stabilized
against both directions. In the same column of \Tref{tab} we also
arrange some suggestive values of the particle masses for $N=2$.
Note that, due to the smallness of $\vcc$, the $\what m_{\bzz\rm
f\ell}$ values are practically equal between each other.

%%%%%%%%%%%%%%%%%%%%%%%%%%%%%%%%%%%%%%%%%%%%%%%%%%%%%%%%%%%%%%%%%%%%
%%%%%%%%%%%%%%%%%%%%%%%%%%%%%%%%%%%%%%%%

\section{\bfseries\scshape Curved Moduli Geometry} \label{cu}

We proceed now to the models with curved internal geometry and
describe below their version for one -- see \Sref{cu1} -- or more
-- see \Sref{cu2} -- moduli.

\subsection{\sc\small\sffamily  Uni-Modular Model} \label{cu1}

The curved moduli geometry is described mainly by the \Ka s
\beq K_{\boldsymbol\pm}=\pm
n\ln\ompm~~\mbox{with}~~\ompm=1\pm\frac{|Z|^2-k^2\zv^4}{n}\label{kpm}
\eeq
and $\zv$ given in \Eref{Zv}. Also $n>0$ and $k$ are real, free
parameters.  The positivity of the argument of logarithm in
\Eref{kpm} implies
\beq  1\pm |Z|^2/n \gtrsim0~~\Rightarrow
\begin{cases}|Z|^2\gtrsim-n&\mbox{for}~~K=\Kp, \\ |Z|\lesssim
\sqrt{n}&\mbox{for}~~K=\Km.
\end{cases}\label{logbound}\eeq
The restriction for $K=\Kp$ is trivially satisfied, whereas this
for $K=\Km$ defines the allowed domain of $Z$ values which lie in
a disc with radius $\sqrt{n}$ and thus, the name disk coordinates.
If we set $k=0$ in \Eref{kpm}, $\Km$ parameterizes \cite{noscale,
su11, nsreview} the coset space $SU(1,1)/U(1)$ whereas $\Kp$ is
associated \cite{su11} with $SU(2)/U(1)$. Thanks to these
symmetries, low $k$ values are totally natural as we explained
below \Eref{sym1}. The K\"ahler metric and (the constant) $\rcal$
in \Eref{rcal} are respectively
\beq\label{ds0} K_{Z\bZ}=\ompm^{-2}~~\mbox{and}~~\rcal_{\rm
\pm}=\pm2/n~~\mbox{for}~~K=K_{\boldsymbol\pm}.\eeq
The last quantity reveals that the \Kmn\ is compact (spherical) or
non-compact (hyperbolic) if $K=\Kp$ or $K=\Km$ respectively. For
this reason, the bold subscripts ${\boldsymbol +}$ or
${\boldsymbol -}$ associated with various quantities below are
referred to $K=\Kp$ or $K=\Km$ respectively.

Repeating the procedure described in \Sref{sup}, we find the form
of $V$ in \Eref{Vsugra}, $V_{\boldsymbol\pm}$, as a function of
$K=\Kpm$ in \Eref{kpm} and $W=\ww$ for $Z=\bz$. This is
\beq V_{\boldsymbol\pm}=\vpm^{\pm n}\ww^2\lf\hspace*{-.1cm}\lf
Z+\vpm\frac{\ww'}{\ww}\rg^2-3\hspace*{-.1cm}\rg~\mbox{with}~~\vpm=1\pm
\frac{Z^2}{n}.\label{Vc0}\eeq
Setting $V_{\boldsymbol\pm}=0$ we see that the corresponding
$\ww=\wwo$ obeys the differential equation
\beq  \frac{d\wwo}{\wwo
}=\frac{\pm\sqrt{3}-Z}{\vpm}dZ.\label{Wdepm}\eeq
This can be resolved yielding two possible forms of $\wwo$,
\beq \wpm^{\pm} = m{v}^{\mp
n/2}_{\boldsymbol\pm}\upm^{\pm1}~~\mbox{for}~~K=\kppm,
\label{w0pm} \eeq
which assure the establishment of Minkowski minima -- cf.
\Eref{wosol}. The corresponding functions $\upm$ can be specified
as follows
\beq \up= e^{\sqrt{3n}\art(Z/n)}~~\mbox{and}~~\um=
e^{\sqrt{3n}\arth(Z/n)}, \label{upm} \eeq
where $\art$ and $\arth$ stand for the functions $\arctan$ and
${\rm arctanh}$ respectively. The superscript $\pm$ in \Eref{w0pm}
correspond to the exponents of $\upm$ and should not be confused
with the bold subscripts ${\boldsymbol \pm}$ with reference to
$\Kpm$.

Combining both Minkowski solutions, $\wpm^{\pm}$ in \Eref{w0pm}
and introducing the shorthand notation -- cf. \Eref{cfpm} --
\beq C_{u\boldsymbol\pm}^{\pm}:=\cp\pm\cm
\upm^{-2}~~\mbox{for}~~K=\kppm,\label{cupm}\eeq
we can obtain the superpotential
\beq \begin{aligned}W_{\Lambda\boldsymbol\pm}&=\cp
W_{0\boldsymbol\pm}^{+}-\cm W_{0\boldsymbol\pm}^{-}\\&= m\vpm^{\mp
n/2}\upm C_{u\boldsymbol\pm}^-
~~\mbox{for}~~K=\kppm\label{Wlpm}\end{aligned}\eeq
which allows for dS/AdS vacua. To verify it, we insert
$K=K_{\boldsymbol\pm}$ and $W=W_{\rm \Lambda\boldsymbol\pm}$ from
\eqs{kpm}{Wlpm} in \Eref{Vsugra} with result
\bea\nonumber \Vpm\hspace*{-.3cm}&&= m^2\ompm^{\pm n}|\vpm|^{\mp
n}|\upm|^2
\Big(-3|\cupmm|^2+n\ompm^2\\\nonumber &&\cdot\left|(\sqrt{3}\cupmp-Z\cupmm)\vpm^{-1}\pm(\bZ-4k^2\zv^3)\cupmm\ompm^{-1}\right|^2\\
&&\cdot\lf\mp|4k^2\zv^3-Z|^2+ n\ompm(1-12k^2\zv^2)\rg^{-1}\Big).
\label{Vpm}\eea
Given that for $\zm=0$ we get $\ompm=\vpm$, we may infer that
$\vev{\Vpm}=\vcc$ shown in \Eref{Vlf} for the directions in
\eqs{vev1}{vev2}. \Eref{ext} for $V=V_{\boldsymbol\pm}$ and
$Z_1:=Z$ is also valid without restrictions for $K=\Kp$ but only
for $n>3$ for $K=\Km$. In fact, employing the decomposition of $Z$
in \Eref{Zzzi} for $\al=1$, we can obtain the scalar spectrum of
our models which includes the sgoldstino components with masses
squared
\beqs\bea \label{mzpm}\what
m_{z\boldsymbol\pm}^2&=&144k^2\mgrpm^2\vev{\vpm^{3/2}\left|{\cupmp}/{\cupmm}\right|}^2;\\
\what
m_{\bzz\boldsymbol\pm}^2&=&4\mgrpm^2\lf1\pm(3/n)\vev{\cupmp/\cupmm}^2\rg\label{mbzpm}\eea\eeqs
for $K=\Kpm$ respectively. The corresponding $\mgr$ according to
\Eref{mgr} -- with $K$ and $W$ given in \eqs{kpm}{Wlpm} -- is
\beq \label{mgrpm}
\mgrpm=m\vev{\upm\cupmm}~~\mbox{for}~~K=\Kpm,\eeq
which may be explicitly written if we use \eqs{upm}{cupm} -- cf.
\Eref{mgrf}. The stability of configurations in \eqs{vev1}{vev2}
is protected for $\what m_{\bzz\boldsymbol-}^2>0$ and $\what
m_{\bzz\boldsymbol-}^2>0$ provided that
\beq
Z<\sqrt{n}~~\mbox{and}~~n>3~~\mbox{for}~~K=\Km.\label{kmbounds}\eeq
Since we expect that $Z\leq1$, the latter restriction is capable
to circumvent both requirements -- see column D in \Tref{tab}.
Inserting the mass spectrum above into the definition of
\Eref{strace}, we can find
\beq{\sf
STr}M_{\boldsymbol\pm}^2=12\mgrpm^2\vev{\cupmp/\cupmm}^2\lf12nk^2\vev{\vpm}^3\pm1\rg.\label{trpm}\eeq
It can be easily verified that the result above is consistent with
the expression of \Eref{strace1} given that $\rcal$ in \Eref{rcal}
is \beq
\vev{\rcal_{\boldsymbol\pm}}=2\lf12k^2n\vev{\vpm}^3\pm1\rg.\label{calrpm}\eeq
%

%%%%%%%%%%%%%%%%%%%%%%%%%%%%%%%%%%%%%%%%%%%%%%%%%%%%%%%%%%%%%%%%%%%%
\begin{figure}[!t]%\vspace*{-.25in}
\includegraphics[width=60mm,angle=-90]{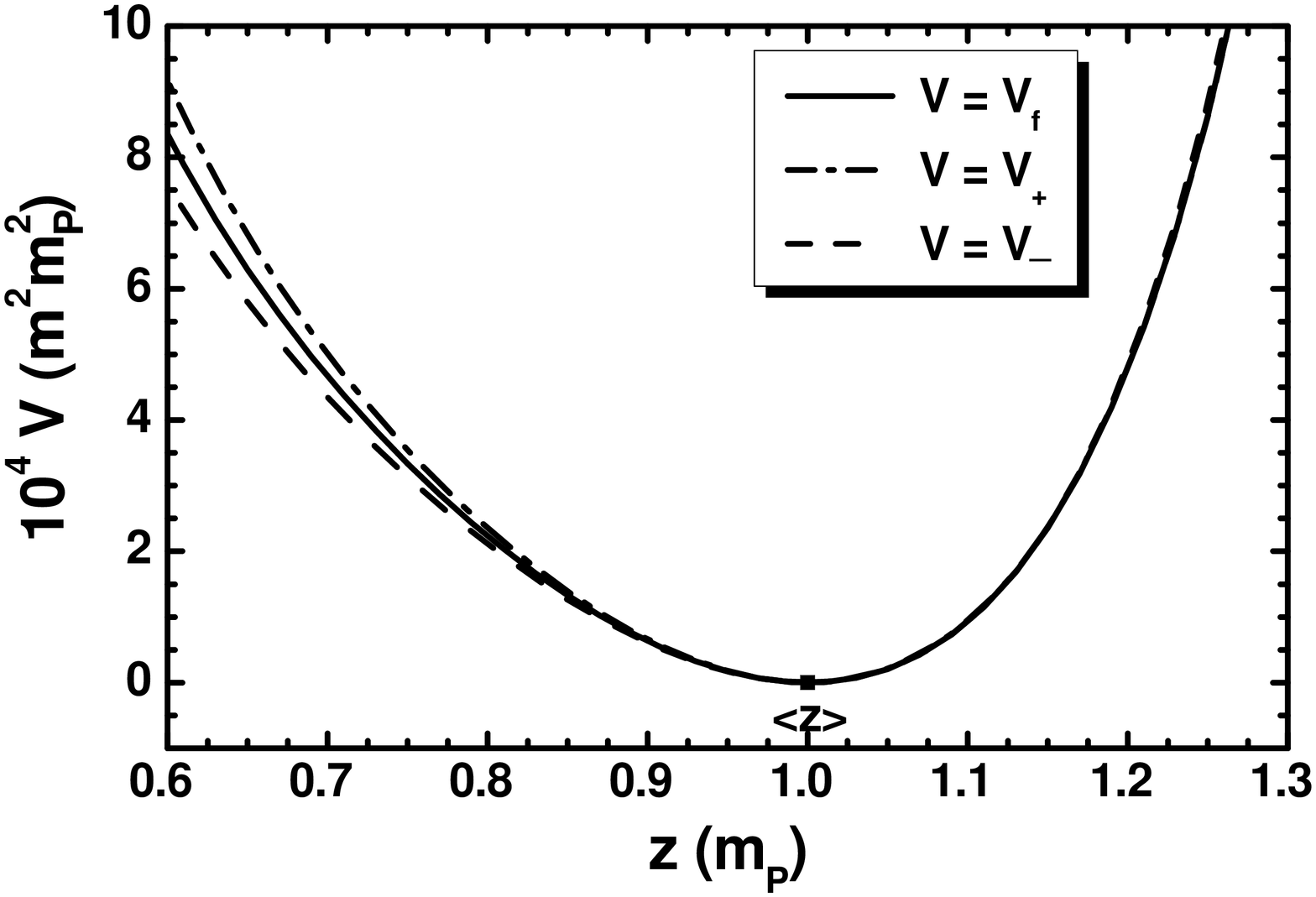}
\caption{\sl \small The (dimensionless) SUGRA potential
$V/m^2\mP^2$ as a function of $z$ for $\bzz=0$ and the settings in
columns A (solid line), C (dot-dashed line) and D (dashed line) of
\Tref{tab}. The value $\vev{z}=\mP$ is also
indicated.}\label{fig3}
\end{figure}
%%%%%%%%%%%%%%%%%%%%%%%%%%%%%%%%%%%%%%%%
%The Minkowski limit of the above results can be easily obtained
%replacing $\cm=0$ or ${\cupmp}/{\cupmm}=1$.

Our analytic results are exemplified in \Fref{fig3}, where we
depict $V_{\boldsymbol+}$ (dot-dashed line) $V_{\boldsymbol-}$
(dashed line) together with $V_{\rm f}$ (solid line) versus $z$
for $\bzz=0$, $k=k_1=0.3$ and the other parameters shown in
columns A, C and D of \Tref{tab}. Note that the selected $n=n_1=4$
for $K=\Km$ protects the stability of the vacuum in \Eref{vev2} as
dictated above. In columns C and D of \Tref{tab} we display some
explicit values of the particle masses encountered for $K=\Km$ and
$\Kp$ respectively. As a consequence of the employed $n$ value in
column D we accidentally obtain $\mgrm=\what m_{\bzz1}=\what
m_{\bzz-}$; ${k_1}$ and $n_1$ obviously coincide with $k$ and $n$
in \Eref{kpm}.

\subsection{\sc\small\sffamily  Multi-Modular Model} \label{cu2}

The generalization of the model above to incorporate more than one
modulus can be performed following the steps of \Sref{fl2}. This
generalization, however, is accompanied by a possible mixing of
the two types of the curved geometry analyzed in \Sref{cu1}. More
specifically, the considered here $K$, includes two sectors with
$\np$ compact components and $\nm$ non-compact ones. It may be
written as
\beq
K_{\np\nm}=\sum_{i=1}^{\np}\ni\ln\omip+\sum_{j=\np+1}^{\nt}\nj\ln\omim,\label{Knt}\eeq
where $\nt=\np+\nm$ and the arguments of the logarithms are
identified as
\beqs\beq \label{omij} \Omega_\al=\begin{cases}
\Omega_{\al\boldsymbol+}&~~\mbox{for}~~\al=i,\\
\Omega_{\al\boldsymbol-}&~~\mbox{for}~~\al=j.\end{cases}\eeq
The symbols $\omipm$ can be collectively defined as
\beq\label{omipm} \omipm=1\pm|Z_\al|^2/n_\al\mp k_\al^2Z_{\rm
v\al}^4\eeq\eeqs
with $Z_{{\rm v}\al}$ is given from \Eref{Zvi} and $\al=i, j$.
When explicitly indicated, summation and multiplication over $i$
and $j$ is applied for the range of their values specified in
\Eref{Knt}. Given that $i$ corresponds to compact geometry
(${\boldsymbol +}$) and $j$ to non-compact (${\boldsymbol -}$) we
remove the relevant indices ${\boldsymbol \pm}$ from the various
quantities to simplify the notation. Under these assumptions, the
positivity of the arguments of $\ln$ implies restrictions only to
$\omim$ -- cf. \Eref{logbound}:
\beq \omim>0~~\Rightarrow~~|\zj|<\sqrt{\nj}\,.
\label{logboundi}\eeq

Along the configurations in \eqs{vevi1}{vevi2} for $\al=i, j$, the
K\"ahler metric is represented by a $\nt\times \nt$ diagonal
matrix
\beq
K_{\al\bbet}=\diag(v_1^{-2},...,v_{N_+}^{-2},v_{N_++1}^{-2},...,v_{\nt}^{-2}),\label{kabpm}\eeq
where we introduce the generalizations of the symbols $\vpm$,
defined in \Eref{Vc0}, as follows
\beq
\vip=(1+\zi^2/\ni)~~\mbox{and}~~\vim=(1-\zj^2/\nj).\label{vipm}\eeq
Also $\rcal$ in \Eref{rcal} includes contributions for both
geometric sectors, i.e,
\beq
\rcal_{\np\nm}=2\mbox{$\sum_i$}\ni^{-1}-2\mbox{$\sum_j$}\nj^{-1}.
\label{rcalpmi}\eeq

Inserting $K=K_{\np\nm}$ from \Eref{Knt} and $W=\wwt(\zi,\zj)$
with $\zi=Z^*_i$ and $\zj=Z^*_j$ in \Eref{Vsugra}, we obtain
\bea \nonumber
V_{N\boldsymbol\pm}&=&\vv^{-2}\Big(\mbox{$\sum_i$}\lf
Z_i\wwt+\partial_i\wwt\vip\rg^2\\&+&\mbox{$\sum_j$}\lf
Z_j\wwt+\partial_j\wwt\vim\rg^2 -3\wwt^2\Big),\label{Vnt}\eea
where the prefactor $\vv$ is defined as follows
\beq \vv=\mbox{$\prod_{i,j}$} \vip^{-\ni/2}\vim^{\nj/2}.
\label{vvdef}\eeq
Setting $V_{N\boldsymbol\pm}=0$ and assuming the ansatz for the
corresponding $\wwt$
\bea \nonumber
&&\wwto(Z_1,...,Z_{\nt})=\prod_{i,j}\wip(\zi)\wim(Z_j)~~\Rightarrow~~\\
&& \frac{\partial_i\wwto}{\wwto}={d\wip\over
d\zi\wip}~~\mbox{and}~~ \frac{\partial_j\wwto}{\wwto}={d\wim\over
dZ_j\wim},\label{Wpmt}\eea
we obtain the separated differential equations
\beq \sum_i\lf \zi+\frac{d\wip}{d\zi\wip}\vip\rg^2+\sum_j\lf
Z_j+\frac{d\wim}{d\zj\wim}\vim\rg^2=3.\eeq
We can solve the equations above if we set
\beq \zi+\frac{d\wip}{d\zi\wip}\vip=|\ai|~~\mbox{and}~~
Z_j+\frac{d\wim}{dZ_j\wim}\vim=|\aj| \eeq
imposing the constraint
\beq \mbox{$\sum_i$}\ai^2+\mbox{$\sum_j$}\aj^2=3, \label{sumai}
\eeq
i.e., the $\ai$ and $\aj$ can be regarded as coordinates of the
hypersphere $\mathbb{S}^{\nt-1}$ with radius $\sqrt{3}$. Solution
of the differential equations above w.r.t $\wip$ and $\wim$ yields
\beq \wip^{\pm}=\vip^{-\ni/2}
\uip^{\pm\ai/\sqrt{3}}~~\mbox{and}~~\wim^{\pm}=\vim^{\ni/2}
\uim^{\pm\aj/\sqrt{3}}\label{Wipm}\eeq
with the generalizations of $\up$ and $\um$ in \Eref{upm} defined
as
\beq \label{upmi} \uip=e^{\sqrt{3\ni}\art(\zi/\sqrt{\ni})}\\
~~\mbox{and}~~\uim=e^{\sqrt{3\nj}\arth(\zj/\sqrt{\nj})}. \eeq
Upon substitution of \Eref{Wipm} into \Eref{Wpmt} we obtain
\beq W_{0N\boldsymbol\pm}^{\pm}=m\prod_{i,j} \wip^{\pm}\wim^{\pm}
=m\vv\uu^{\pm1}, \label{Wnt0}\eeq
where we define the function
\beq \uu=\mbox{$\prod_{i,j}$}
\uip^{\ai/\sqrt{3}}\uim^{\aj/\sqrt{3}}. \label{uu}\eeq
Introducing the generalized $C$ symbols -- cf. \Eref{cfpm} --
\beq \Cupm:=\cp\pm\cm \uu^{-2},\label{cpmi}\eeq
we combine both solutions in \Eref{Wnt0} as follows
\bea W_{\Lambda N\boldsymbol\pm}=\cp W_{0N\boldsymbol\pm}^{+}-\cm
W_{0N\boldsymbol\pm}^{-}= m\vv \uu \Cum.\label{Wtll}\eea

Plugging $K=K_{\np\nm}$ and $W=W_{\Lambda N\boldsymbol\pm}$ from
\eqs{Knt}{Wtll} in \Eref{Vsugra} we find
\beq\begin{aligned} V_{N\boldsymbol\pm}&=
m^2\lf\mbox{$\prod_{ij}$}\omip^{\ni}\omim^{-\nj}\rg|\vv\uu|^2
\Bigg(-3|\Cum|^2\\
&+\sum_i\Big(\ni\omip^2\cdot\left|(\ai\Cup-\zi\Cum)\vip^{-1}\right.
\\&+\left.(Z^*_i-4k_i^2\zvi^3)\Cum\omip^{-1}\right|^2\\
&\cdot\lf-\left|4\ki^2\zvi^3-\zi\right|^2+\ni\omip\lf1-12\ki^2\zvi^2\rg\rg^{-1}\Big)
\\&+\sum_j\Big(\nj\omim^2\cdot\left|(\aj\Cup-\zj\Cum)\vim^{-1}\right.
\\&-\left.(Z^*_j-4k_j^2\zvj^3)\Cum\omim^{-1}\right|^2\\
&\cdot\lf\left|4\kj^2\zvj^3-\zj\right|^2+\nj\omim\lf1-12\kj^2\zvj^2\rg\rg^{-1}\Big)
\Bigg).
\end{aligned}\label{VNpm}\eeq
Note that there are slight differences between the terms with
subscripts $i$ and $j$ due to our convention in \Eref{omij} -- cf.
\Eref{Vpm}. The settings in \eqs{vevi1}{vevi2} consist honest
dS/AdS vacua since $\vev{V_{N\boldsymbol\pm}}=\vcc$ given in
\Eref{Vlf}. However, the conditions in \Eref{ext} for
$V=V_{N\boldsymbol\pm}$ and $\al=i, j$ are met only after imposing
upper bound on $\vrm_j$ and $\aj$. To determine this, we extract
the masses squared of the $2\nt$ scalar components of $\zi$ and
$\zj$ in \Eref{Zzzi} which are
\beqs \bea \what
m_{zi\boldsymbol+}^2&=&48\ki^2\ai^2\mgru^2\vev{\vip^{3/2}{\Cup}/{\Cum}}^2;\label{mzip}\\
\what
m_{zj\boldsymbol-}^2&=&48k_j^2\aj^2\mgru^2\vev{\vim^{3/2}{\Cup}/{\Cum}}^2;\label{mzim}\\
\what m_{\bzz i\boldsymbol+}^2&\simeq&4\mgru^2\lf1+\ai^2/\ni\rg;\label{mbzip}\\
\what m_{\bzz
j\boldsymbol-}^2&\simeq&4\mgru^2\lf1-\aj^2/\nj\rg,\label{mbzim}\eea\eeqs
where we restore the $\boldsymbol\pm$ symbols for clarity and we
neglect for simplicity terms of order $(\cm)^2$ in the two last
expressions. We also compute $\mgr$ upon substitution of
\eqs{Knt}{Wtll} into \Eref{mgr} with result
\beq \label{mgrpmN} \mgru=m\vev{\uu\Cum}.\eeq
As in the case of \Sref{fl2}, the relevant matrix $M_0^2$ in
\Eref{mbos} turns out to be essentially diagonal since the
non-zero elements appearing in the $\bzz_\al-\bzz_{\beta}$
positions with $\al,\beta=i,j$ are proportional to $\vcc$ and can
be safely ignored compared to the diagonal terms. From
\eqs{mzim}{mbzim}, we notice that positivity of $\what m_{zj-}^2$
and $\what m_{\bzz j-}^2$ dictates
\beq
Z_j<\sqrt{\nj}~~\mbox{and}~~|\aj|<\sqrt{\nj}.\label{ajdom}\eeq
These restrictions together with \Eref{sumai} delineate the
allowed ranges of parameters in the hyperbolic sector. We also
obtain $\nt-1$ Weyl fermions with masses $\what m_{\tilde z\tilde
\al}=\mgr$ with $\tilde \al=1,...,\nt-1$. Inserting the mass
spectrum above into \Eref{strace} we find
\beq\begin{aligned}{\sf
STr}M_{N\boldsymbol\pm}^2&\simeq6\mgru^2\Bigg((\nt-1)+\frac23\lf\mbox{$\sum_i$}\frac{\ai^2}{\ni}+
\mbox{$\sum_j$}\frac{\aj^2}{\nj}\rg+\\&
8\vev{\frac{\Cup}{\Cum}}^2\lf\mbox{$\sum_i$}\ai^2\ki^2\vev{\vip}^3+
\mbox{$\sum_j$}\aj^2\kj^2\vev{\vim}^3\rg\Bigg).\end{aligned}\label{trNpm}\eeq
It can be checked that this result is consistent with
\Eref{strace2}.

%%%%%%%%%%%%%%%%%%%%%%%%%%%%%%%%%%%%%%%%%%%%%%%%%%%%%%%%%%%%%%%%%%%%
\begin{figure}[t]%
\epsfig{file=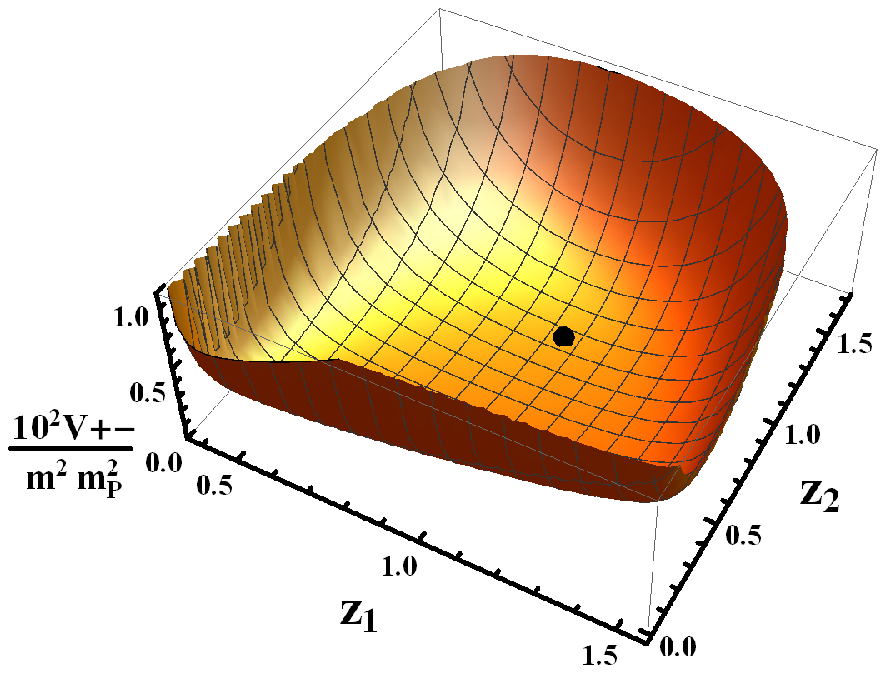,width=8.7cm,,angle=-0}
\vspace*{2.3in} \caption{\sl \small The (dimensionless) SUGRA
potential $V_{\boldsymbol+-}/m^2 \mP^2$ in \Eref{VNpm} as a
function of $z_1$ and $z_2$ in \Eref{Zzzi} for $\bzz_1=\bzz_2=0$
and the inputs shown in column E of \Tref{tab}. The location of
the dS vacuum in \Eref{vevi2} is also depicted by a thick black
point.}\label{fig4}
\end{figure}
%%%%%%%%%%%%%%%%%%%%%%%%%%%%%%%%%%%%%%%%

For $N=2$, $\nm=\np=1$ and the parameters shown in column E of
\Tref{tab}, we present in \Fref{fig4} the relevant
$V_{N\boldsymbol\pm}$, $V_{2\boldsymbol\pm}=V_{\boldsymbol+-}$ --
conveniently normalized -- versus $z_1$ and $z_2$ in \Eref{Zzzi}
fixing $\bzz_1=\bzz_2=0$. It is clearly shown that the vacuum of
\Eref{vevi2}, depicted by a bold point, is indeed stable. In
column E of \Tref{tab} we arrange also some representative masses
(in \GeV) of the particle spectrum for $\nt=2$. From the
parameters listed there we infer that $a_2=\sqrt{2}<\sqrt{n_2}=2$
and so \eqs{sumai}{ajdom} are met.

\section{\bfseries\scshape Generalization}\label{mx}

It is certainly impressive that the models described in
Sec.~\ref{fl2} and \ref{cu2} can be combined in a simple and
(therefore) elegant way. We here just specify the utilized $K$ and
$W$ of a such model and restrict ourselves to the verification of
the results. In particular, we consider the following $K$
\beq K_{N\rm f\boldsymbol\pm}=K_{N\rm
f}+K_{\np\nm},\label{Knfpm}\eeq
which incorporates the individual contributions from
\eqs{Kfn}{Knt}. It is intuitively expected that the required $W$
for achieving dS/AdS vacua has the form -- cf.~\eqs{Wll}{Wtll}
\beq W_{\Lambda N\rm f\boldsymbol\pm}=\cp  W_{0N\rm
f\boldsymbol\pm}^{+}+\cm W_{0N\rm
f\boldsymbol\pm}^{-},\label{Wnfpm}\eeq
where the definitions of $W_{0N\rm f\boldsymbol\pm}^{\pm}$ follow
those in \eqs{Wnf0}{Wnt0} respectively. Namely, we set
\beq W_{0N\rm
f\boldsymbol\pm}^{\pm}=m\calw\vv(F\uu)^{\pm1},\label{Wfu}\eeq
where the parameters $\all,\ai$ and $\aj$, which enter the
expressions of the functions $F$ and $\uu$ in \eqs{wfi}{uu},
satisfy the constraint -- cf. \eqs{aif}{sumai}
\beq
\mbox{$\sum_\ell$}\all^2+\mbox{$\sum_i$}\ai^2+\mbox{$\sum_j$}\aj^2=3.
\label{sumail} \eeq
I.e., they lie at the hypersphere $\mathbb{S}^{\ntt-1}$ with
radius $\sqrt{3}$ and $\ntt=\nf+\nt$. If we introduce, in
addition, the $C$ symbols -- cf.~\eqs{cfpmi}{cpmi} --
\beq \Cfupm:=\cp\pm\cm(F\uu)^{-2},\label{cfu}\eeq
$W_{\Lambda N\rm f\boldsymbol\pm}$ in \Eref{Wnfpm} is simplified
as
\beq W_{\Lambda N\rm f\boldsymbol\pm}=m\calw \vv F\uu
\Cfum.\label{Wnfpm1} \eeq

Plugging $K=K_{N\rm f\boldsymbol\pm}$ and $W=W_{\Lambda N\rm
f\boldsymbol\pm}$ from \eqs{Knfpm}{Wnfpm1} into \Eref{Vsugra} we
obtain
\bea V_{N\rm f\boldsymbol\pm}&=& m^2e^{K_{N\rm
f}}\lf\mbox{$\prod_{ij}$}\omip^{\ni}\omim^{-\nj}\rg|F\vv\uu|^2
\Bigg(-3|\Cfum|^2\nonumber\\
&+&\sum_i\Big(\ni\omip^2\cdot\left|(\ai\Cfup-\zi\Cfum)\vip^{-1}\right.
\nonumber\\&+&\left.(Z^*_i-4k_i^2\zvi^3)\Cfum\omip^{-1}\right|^2\nonumber\\
&\cdot&\lf-\left|4\ki^2\zvi^3-\zi\right|^2+\ni\omip\lf1-12\ki^2\zvi^2\rg\rg^{-1}\Big)\nonumber
\\&+&\sum_j\Big(\nj\omim^2\cdot\left|(\aj\Cfup-\zj\Cfum)\vim^{-1}\right.\nonumber
\\&-&\left.(Z^*_j-4k_j^2\zvj^3)\Cfum\omim^{-1}\right|^2\nonumber\\
&\cdot&\lf\left|4\kj^2\zvj^3-\zj\right|^2+\nj\omim\lf1-12\kj^2\zvj^2\rg\rg^{-1}\Big)\nonumber
\\&+&\sum_\ell\left|\all\Cfup-\zml\Cfum-4\kl^2\zvl^3\Cfum\right|^2\nonumber\\
&\cdot&\lf1-12\kl^2\zvl^2\rg^{-1} \Bigg).\label{VNfpm}\eea
Once again, we infer that \eqs{vevi1}{vevi2} consist dS/AdS vacua
since $\vev{V_{N\rm f\boldsymbol\pm}}=\vcc$ -- see \Eref{Vlf} --
and \Eref{ext} with $V=V_{N\rm f\boldsymbol\pm}$ and $\al=\ell, i,
j$ is fulfilled if we take into account the restrictions in
\eqs{ajdom}{logboundi}.

The \Gr\ mass is derived from \Eref{mgr}, after substituting $K$
and $W$ from \eqs{Knfpm}{Wnfpm1} respectively. The result is
\beq \label{mgrpmNf} \mgrfu=m\vev{F\uu\Cfum}.\eeq
From \eqs{mbos}{mfer} with $\al=\ell, i, j$, we can obtain the
mass spectrum of the present model which includes $2\ntt$ real
scalars and $\ntt-1$ Weyl fermions with masses $\what m_{\tilde
z\al}=\mgr$ where $\al=1,...,\ntt-1$. The masses squared of the
$2\nf$ scalars are given in the $\cm\to0$ limit by \eqs{mzi}{mbzi}
for $\Cfp/\Cfm=1$ and $\mgrnf$ replaced by $\mgrfu$. In the same
limit the masses squared of the $2\nt$ scalars are given by
Eq.~(\ref{mzip}) -- (\ref{mbzim}) for $\Cup/\Cum=1$ and $\mgru$
replaced by $\mgrfu$.

\begin{figure}[t]%
\epsfig{file=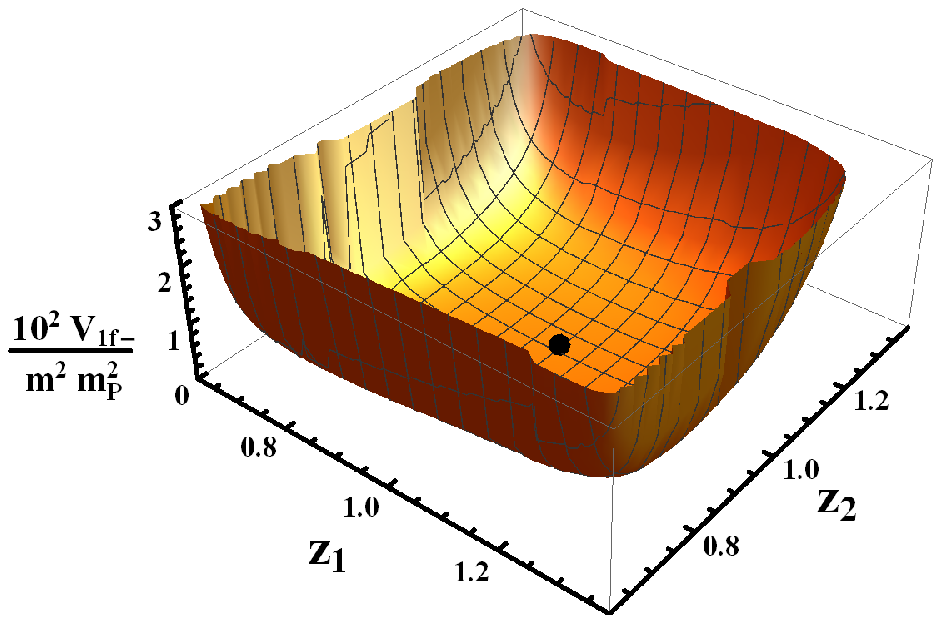,width=8.7cm,,angle=-0}
\vspace*{2.3in} \caption{\sl \small The (dimensionless) SUGRA
potential $V_{\rm f\boldsymbol-}/m^2 \mP^2$ in \Eref{VNfpm} as a
function of $z_1$ and $z_2$ for $\bzz_1=\bzz_2=0$ -- see
\Eref{Zzzi} -- and the inputs shown in column F of \Tref{tab}. The
location of the dS vacuum in \Eref{vevi2} is also depicted by a
thick black point.}\label{fig5}
\end{figure}

To provide a pictorial verification of our present setting, we
demonstrate in \Fref{fig5} the three-dimensional plot of $V_{N\rm
f\boldsymbol\pm}$ with $\nf=\nm=1$ and $\np=0$, i.e. $V_{1\rm
f\boldsymbol-}$, versus $z_1$ and $z_2$ for $\bzz_1=\bzz_2=0$ --
see \Eref{Zzzi} -- and the other parameters arranged in column F
of \Tref{tab}. Note that the subscripts $1$ and $2$ of $z$
correspond to $\ell=1$ and $j=1$ and the validity of
\eqs{ajdom}{sumail} is protected. It is evident that the ground
state, depicted by a tick black point is totally stable. Some
characteristic values of the masses of the relevant particles are
also arranged in column F of \Tref{tab}.

%%%%%%%%%%%%%%%%%%%%%%%%%%%%%%%%%%%%%%%%%%%%%%%%%%%%%%%%%%%%%%%%%%%%
%%%%%%%%%%%%%%%%%%%%%%%%%%%%%%%%%%%%%%%%

\section{\bfseries\scshape Link to the Observable Sector}\label{obs}

Our next task is to study the transmission of the SUSY breaking to
the visible world.  Here we restrict for simplicity ourselves to
the cases with just one Goldstino superfield, $Z$. To implement
our analysis, we introduce the chiral superfields of the
observable sector $\phc_\al$ with $\al=1,...,5$ and assume the
following structure -- cf.~\cref{nilles,susyr,soft} -- for the
total superpotential, $W_{\rm HO}$, of the theory
\beq \label{Who} W_{\rm HO}=\whi(Z) + W_{\rm O}\lf\phc_\al\rg,\eeq
where $\whi$ is given by \Eref{Wl} or \Eref{Wlpm} for flat or
curved $Z$ geometry respectively whereas  $W_{\rm O}$ has the
following generic form
\beq W_{\rm O}=h \phc_1\phc_2\phc_3+\mu\phc_4\phc_5.
\label{w0}\eeq
with $h$ and $\mu$ free parameters. On the other hand, we consider
three variants of the total $K$ of the theory, $K_{\rm HO}$,
ensuring universal SSB parameters for $\phc_\al$:
\beqs\bel
K_{1\rm HO}&=\kh(Z)+\mbox{$\sum_\al$}|\phc_\al|^2;~~~\label{K1}\\
K_{2\rm HO}&=\kh(Z)+N_{\rm
O}\ln\left(1+\mbox{$\sum_\al$}|\phc_\al|^2/N_{\rm
O}\right),\label{K2} \end{align}
where $\kh(Z)$ may be identified with $K_{\rm f}$ in \Eref{Kf} or
$K_{\pm}$ in \Eref{kpm} for flat or curved $Z$ geometry
respectively whereas $N_{\rm O}$ may remain unspecified. For
curved $Z$ geometry we may introduce one more variant
\beq K_{3\rm HO}=\pm
n\ln\Big(\ompm\pm\mbox{$\sum_\al$}|\phc_\al|^2/n\Big).\label{K3}\eeq\eeqs
If we expand the $K_{\rm HO}$'s above for low $\phc_\al$ values,
these may assume the form
\beqs\beq \label{Kho} K_{\rm HO}=\kh(Z)+\wkhi(Z)|\phc_\al|^2,\eeq
with $\wkhi$ being identified as
\beq \label{wk} \wkhi=\begin{cases} 1&\mbox{for}~~K_{\rm
HO}=K_{1\rm HO}, K_{2\rm HO};\\\ompm^{-1}&\mbox{for}~~K_{\rm
HO}=K_{3\rm HO}\,.\end{cases}\eeq\eeqs

Adapting the general formulae of \cref{soft, susyr} to the case
with one hidden-sector field and tiny $\vev{V}$, we obtain the SSB
terms in the effective low energy potential which can be written
as
\beq V_{\rm SSB}= m_\al^2 |\what\phc_\al|^2+\lf A\what h
\what\phc_1\what\phc_2\what\phc_3+ B\what \mu
\what\phc_4\what\phc_5+{\rm h.c.}\rg, \label{vsoft} \eeq
where the rescaled parameters are
\beq \what h=e^{\veva{K_{\rm H}}/2}\veva{\wtilde K_{\rm H}}^{-3/2}
h~~\mbox{and}~~\what\mu=e^{\veva{K_{\rm H}}/2}\veva{\wtilde K_{\rm
H}}^{-1}\mu\eeq
and the canonically normalized fields are
$\what\phc_\al=\veva{\wtilde K_{\rm H}}^{1/2}\phc_\al$.

In deriving the values of the SSB parameters above, we distinguish
the cases:

\subparagraph{\small\sf (a)} For flat $Z$ geometry, i.e.
$\kh=K_{\rm f}$, we see from \Eref{wk} that $\wkhi$ is constant
for both adopted $K_{\rm HO}$'s and so, the results are common.
Substituting
\beq\vev{F^Z}=\sqrt{3}\mgrf~~\mbox{and}~~\vev{\partial_Z\kh}=\vrm,
\label{aux1f}\eeq
into the relevant expressions \cite{susyr} we arrive at
\beq\begin{aligned} \label{ssbf} m_\al &=\lf1+e^{\sqrt{6}\vrm}{\vcc}/{m^2\vev{\cfm}^2}\rg\mgrf\simeq\mgrf,\\
A &=-\sqrt{\frac{3}{2}}\vrm\vev{\frac{\cfp}{\cfm}}\mgrf\simeq\sqrt{\frac{3}{2}}\vrm\mgrf,\\
B&= A-\mgrf,
\end{aligned}\eeq
%\simeqe^{\vrm^2/2}\lf\sqrt{3}\vrm-1\rg\mgrf
where $(\what h, \what\mu) = e^{\vrm^2/4} (h,\mu)$ and the last
simplified expressions are obtained in the realistic limit
$\cm\to0$ which implies $\cfp/\cfm\to-1$.

\subparagraph{\small\sf (b)} For curved $Z$ geometry, i.e.
$\kh=\Kpm$, we can distinguish two subcases depending on which
$K_{\rm HO}$ from those shown in \Eref{K1} -- (\ref{K3}) is
selected. Namely,

\begin{itemize}

\item If $K_{\rm HO}=K_{1\rm HO}$ or $K_{2\rm HO}$, then $\wkhi$
in \Eref{wk} is $Z$ independent. For $\kh=\Kpm$ respectively we
find
\begin{align}\label{ssbpm1} (m_\al, A, B)
=\lf1,\sqrt{3/2}\vrm,A/\mgrpm-1\rg\mgrpm, \end{align}
where $(\what h, \what\mu) = \vev{\vpm}^{\pm n/2} (h,\mu)$ and we
take into account the following
$$\vev{F^Z}=\sqrt{3}\vev{\vpm}\mgrpm~~\mbox{and}~~
~\vev{\partial_Z\Kpm}=\vev{Z\vpm^{-1}}.\hspace*{-0.4cm}$$

\item If $K_{\rm HO}=K_{3\rm HO}$, then $\wkhi$ in \Eref{wk} is
$Z$ dependent. Inserting the expressions
%={3}\vev{\partial_Z\ln\wkhi^3}/2
$$\vev{\partial_Z\ln\wkhi}=\frac1{n\vev{\vpm}^2}~~\mbox{and}
\vev{\partial_Z\ln\wkhi^2}=\frac{2\vrm}{n\vev{\vpm}}\hspace*{-0.4cm}
$$
into the general formulae \cite{soft, susyr} we end up with the
following results for $\kh=\Kpm$ correspondingly:
\beq\begin{aligned} \label{ssbpm2} m_\al &=\sqrt{1\pm3/n}\,\mgrpm;\\
A &=\sqrt{3/2}\vrm(1\pm3/n)\mgrpm;\\
B&=\lf\sqrt{3/2}\vrm(1\pm2/n)-1\rg\mgrpm,\hspace*{-0.7cm}
\end{aligned}\eeq
where $\what h= \vev{\vpm}^{(3\pm n)/2} h$ and $\what\mu =
\vev{\vpm}^{(2\pm n)/2}\mu$.
%\label{Apm2}\label{Bpm2}
\end{itemize}

In both cases above we take $\cm\simeq0$ for simplicity. Note that
the condition $n>3$ for $K=\Km$ which is imperative for the
stability of the configurations in \eqs{vev1}{vev2} -- see
\Eref{mbzpm} -- implies non-vanishing SSB parameters too. Taking
advantage from the numerical inputs listed in columns A, C and D
of \Tref{tab} (for the three unimodular models) we can obtain some
explicit values for the SSB parameters derived above -- restoring
units for convenience. Our outputs are arranged in the three
rightmost columns of \Tref{tabs} for the specific forms of $\whi,
\kh$ and $K_{\rm HO}$ in \eqs{Who}{Kho} shown in the three
leftmost columns.  We remark that there is a variation of the
achieved values of SSB parameters which remain of the order of the
$\Gr$ mass in all cases.

%Evidently, $\wkhi=1$ corresponds to $;ljkklK_{\rm HO}=K_{1,2\rm HO}$
%whereas $\wkhi=1/\ompm$ corresponds to $K_{\rm HO}=K_{3\rm HO}$.

%%%%%%%%%%%%%%%%%%%%%%%%%%%%%%%%%%%%%%%%%%%%%%%%%%%%%%%%%%%%%%%%%%%
\begin{table}[t!]
\caption{\sl SSB Parameters: a Case Study for the Inputs of
\Tref{tab}.}
%\begin{ruledtabular}
\begin{tabular}{c@{\hspace{0.3cm}}|@{\hspace{0.3cm}}c@{\hspace{0.3cm}}
|@{\hspace{0.3cm}}c@{\hspace{0.3cm}}||c@{\hspace{0.5cm}}|@{\hspace{0.5cm}}c@{\hspace{0.5cm}}
|@{\hspace{0.5cm}}c@{\hspace{0.5cm}}}\toprule
\multicolumn{3}{c||}{\sc Input Settings}&\multicolumn{3}{c}{\sc
SSB Parameters in $\GeV$}\\ \hline
$\whi$&$\kh$&$\wkhi$&{\hspace{0.3cm}}$m_\al$&$|A|$&$|B|$\\
\hline\hline
$W_{\Lambda\rm f}$&$K_{\rm
f}$&$1$&{\hspace{0.3cm}}$170$&$208$&$378$\\\hline
$W_{\Lambda\boldsymbol+}$&$K_{\rm \boldsymbol+}$&$1$&{\hspace{0.3cm}}$145$&$177$&$32$\\
$W_{\Lambda\boldsymbol+}$&$K_{\rm
\boldsymbol+}$&$1/\omp$&{\hspace{0.3cm}}$290$&$711$&$388$\\\hline
$W_{\Lambda\boldsymbol-}$&$K_{\rm \boldsymbol-}$&$1$&{\hspace{0.3cm}}$179$&$220$&$40$\\
$W_{\Lambda\boldsymbol-}$&$K_{\rm
\boldsymbol-}$&$1/\omm$&{\hspace{0.3cm}}$90$&$55$&$69$\\
\botrule
\end{tabular}\label{tabs}
%\end{ruledtabular}
\end{table}

\section{\bfseries\scshape Conclusions} \label{con}

We have extended the approach of \cref{noscale18,noscale19},
proposing new no-scale SUGRA models which lead to Minkowski, dS
and AdS vacua without need for any external uplifting mechanism.
We first provided a simple but general enough argument which
assists to appreciate the effectiveness of our paradigm. We then
adopted specific single-field models and showed that the
achievement of dS/AdS solutions using pairs of Minkowski ones
works perfectly well for flat -- see \eqs{Kf}{Wl} -- and
hyperbolic or spherical geometry -- see \eqs{kpm}{Wlpm}. We also
broadened these constructions to multi-field models -- see
Sec.~\ref{fl2}, \ref{cu2} and \ref{mx}. Within each case we
derived the SUGRA potential and the relevant mass spectrum paying
special attention to the stability of the proposed solutions.
Typical representatives of our results were illustrated in
Fig.~\ref{fig1} -- \ref{fig5} employing numerical inputs from
\Tref{tab}. We provided, finally, the set of the soft
SUSY-breaking parameters induced by our unimodular models linking
them to a generic observable sector -- see Eqs. ~(\ref{ssbf}) -
(\ref{ssbpm2}). We verified -- see \Tref{tabs} -- that their
magnitude is of the order of the gravitino mass.

As stressed in \cref{noscale19, noscaleinfl}, this kind of
constructions, based exclusively in SUGRA, can be considered as
part of an effective theory valid below $\mP$. However, the
correspondence between K\"ahler and super-potentials which yields
naturally Minkowski, dS and AdS (locally stable) vacua with broken
SUSY may be a very helpful guide for string theory so as to
establish new possible models with viable low energy
phenomenology. As regards the ultraviolet completion, it would be
interesting to investigate if our models belong to the string
landscape or swampland \cite{vafa}. Note that the swampland string
conjectures are generically not satisfied in SUGRA-based models
but there are suggestions \cite{ferara, sevilla} which may work in
our framework too. One more open issue is the interface of our
settings with inflation. We aspire to return on this topic soon
taking advantage from other similar studies \cite{noscaleinfl,
ketov, lhclinde, ant} -- see \cref{deinf}. At last but not least,
let us mention that the achievement of the present value of the
dark-energy density parameter in \Eref{omde} requires an inelegant
fine tuning, which may be somehow alleviated if we take into
account contributions from the electroweak symmetry breaking
and/or the confinement in quantum chromodynamics \cite{noscale19,
noscaleinfl}.

Despite the shortcomings above, we believe that the establishment
of novel models for SUSY breaking with a natural emergence of
Minkowski and dS/AdS vacua can be considered as an important
development which offers the opportunity for further explorations
towards several cosmo-phenomenological directions.

%\newpage

\paragraph*{\small\bfseries\scshape Acknowledgment} {\small This research work
was supported by the Hellenic Foundation for Research and
Innovation (H.F.R.I.) under the ``First Call for H.F.R.I. Research
Projects to support Faculty members and Researchers and the
procurement of high-cost research equipment grant'' (Project
Number: 2251).}

\paragraph*{\small\bfseries\scshape Dedication} {\small  I would like
to dedicate the present paper to the memory of T.~Tomaras, an
excellent University teacher who let his imprint on my first
post-graduate steps.}

\appendix

\renewenvironment{subequations}{%
\refstepcounter{equation}%
% \theparentequation{\theequation}%
\setcounter{parentequation}{\value{equation}}%
  \setcounter{equation}{0}
  \def\theequation{A\theparentequation{\ftn\sffamily\alph{equation}}}%
  \ignorespaces
}{%
  \setcounter{equation}{\value{parentequation}}%
  \ignorespacesafterend
}
\renewcommand{\thesubsection}{{\small\sf\arabic{subsection}}}

\section{\bfseries\scshape Mass Formulae In SUGRA}\label{app1}

We here generalize our formulae in \cref{susyr} for $N$ chiral
multiplets and dS/AdS vacua. Let us initially remind that central
role in the SUGRA formalism plays the K\"ahler-invariant function
expressed in terms of the \Ka\ $K$ and the superpotential $W$ as
follows
\beq \label{Gdef} G := K + \ln |W|^2. \eeq
Using it we can derive the F-term scalar potential \cite{nilles}
\beq \label{Vsugra} V = e^{G}\lf G^{\al\bbet} G_\al
G_\bbet-3\rg=e^{K}|W|^2\lf K^{\al\bbet} G_\al G_\bbet-3\rg, \eeq
where the subscripts of quantities $G$ and $K$ denote
differentiation w.r.t the superfields $Z_\al$ and
$G^{\al\bbet}=K^{\al\bbet}$ is the inverse of the \Kaa\ metric
$K_{\al\bbet}$. The spontaneous SUSY breaking takes place
typically at a (locally stable) vacuum or flat direction of $V$
which satisfies the extremum and minimum conditions
\beq\vev{\partial_{\al}V}=\vev{\partial_{\bar
\al}V}=0~~\mbox{and}~~\what m_A^2>0. \label{ext}\eeq
Here $\partial_{\al}:=\partial/
\partial Z_\al$ and $\partial_{\bar\al}:=\partial/
\partial Z^*_{\bar\al}$ with the scalar components
of the superfields denoted by the same superfield symbol. Also
$\what m_A^2$ are the eigenvalues of the $2N\times2N$ mass-squared
matrix $M^2_0$ of the (canonically normalized) scalar fields which
is computed applying the formula
\beq  \label{mbos} M^2_0= \mtt{\vev{\partial_{\what
\al}\partial_{\what \bt}V}}{\vev{\partial_{\what
\al}\partial_{\what \bbet}V}}{\vev{\partial_{\what
\bbet}\partial_{\what \al}V}}{\vev{\partial_{\what
\bbet}\partial_{\what \aal}V}}, \eeq
where $\partial_{\what A}:=\partial/
\partial \what Z_A$ with $A=\al$ or $\bar \al$ and $\what Z_\al=\sqrt{K_{\al\aal}}Z_\al$ given
that the $K$'s considered in our work are diagonal.

The aforementioned $M^2_0$ is one of the mass-squared matrices
$M_J^2$ of the particles with spin $J$, composing the spectrum of
the model. They obey the super-trace formula \cite{nilles, mass}
\beqs\bel\label{strace} {\sf STr}M^2&:=\sum_{J=0}^{3/2}(-1)^{2J}
(2J+1)\Tr
M_J^2\\\nonumber&=2\mgr^2\Big(\vev{(N-1)(1+V/\mgr^2)}\\&+\vev{G_\al
G^{\al\bbet}R_{\beta\bar
\gamma}G^{\gamma\bar\delta}G_{\bar\delta}}\Big),\label{strace2}\end{align}\eeqs
where $R_{\al\bbet}$ is the (moduli-space) Ricci curvature which
reads
\beq \label{ricci} R_{\al\bbet}=-\partial_\al\partial_\bbet\ln
{\rm det}\lf G_{\gamma\bar\delta}\rg.\eeq
Note that \Eref{strace2} provides a geometric computation of ${\sf
STr}M^2$ which can be employed as an consistency check for the
correctness of a direct computation via the extraction of the
particle spectrum by applying \Eref{strace}.

The factor $N-1$ in the first term of \Eref{strace2} reflects the
fact that we obtain one fermion with spin $1/2$ less than the
number $N$ of the chiral multiplets. This is because one such
fermion, known as goldstino, is absorbed by the gravitino ($\Gr$)
with spin $3/2$ according to the super-Higgs mechanism
\cite{nilles}. The $\Gr$ mass squared is evaluated as follows
\beq \label{mgr} \mgr^2=\vev{e^{G}}=\frac13\vev{G_{\al\bbet}F^\al
F^{*\bbet}-V},\eeq
where the F terms are defined as \cite{soft}
\beq \label{Fz} F^\al:= e^{G/2}K^{\al\bbet}G_\bbet
~~\mbox{and}~~F^{*\aal}:= e^{G/2}K^{\aal\bt}G_\bt\,.\eeq

In our work we compute also the elements of $M_{1/2}$, i.e., the
masses of the (canonically normalized) chiral fermions, $\tilde
Z_\al,$ which can be found applying the formula
\beq \label{mfer}
m_{\al\bt}=\mgr\vev{G_{\al\bt}+(1-2/U)G_{\al}G_{\bt}}\vev{G_{\al\aal}G_{\bt\bbet}}^{-1/2},\eeq
where $G_{\al\bt}$ is defined in terms of the K\"{a}hler-covariant
derivative $D_\al$ as
\beq \label{Gab} G_{\al\bt}:=D_\al G_\bt=\partial_\al
G_{\bt}-\Gamma^\gamma_{\al\bt}G_\gamma,\eeq
with $\Gamma^\gamma_{\al\bt}=K^{\gamma\bar\gamma}\partial_\al
K_{\bt\bar\gamma}$ and $U$ takes into account a possible
non-vanishing $\vev{V}$, i.e.,
\beq U = 3+\vev{V}/\mgr^2.\eeq
In \Eref{mfer} care is taken so as to canonically normalize the
various fields and remove the mass mixing between $\Gr$ and fields
with spin $1/2$ in the SUGRA lagrangian.

Let us, finally, note that \Eref{strace2} can be significantly
simplified for $N=1$ since it can be brought into the form
\bea \nonumber {\sf
STr}M^2&=2\mgr^2\veva{G_{Z\bz}^{-2}G_ZG_{\bz}R_{Z\bz}}\\&=2\mgr^2\vev{(3+V/\mgr^2){\mathcal
R}}, \label{strace1}\eea
where we make use of \Eref{Vsugra} and the definition of the
scalar curvature $\rcal$ which is
\beq \label{rcal} \rcal=G^{\al\bbet}R_{\al\bbet}.\eeq
Note that the first term of \Eref{strace2} vanishes for $N=1$ due
to the super-Higgs effect.

%In order to sum up the masses of fields for each spin separately,
%such a mixing term should be removed by diagonalizing the fermions

\section{\bfseries\scshape Half-Plane Parametrization}\label{hp}

\renewenvironment{subequations}{%
\refstepcounter{equation}%
% \theparentequation{\theequation}%
\setcounter{parentequation}{\value{equation}}%
  \setcounter{equation}{0}
  \def\theequation{B\theparentequation{\ftn\sffamily\alph{equation}}}%
  \ignorespaces
}{%
  \setcounter{equation}{\value{parentequation}}%
  \ignorespacesafterend
}

In this Appendix we employ the half-plane parametrization of the
hyperbolic geometry which allows us to compare our results in
\Sref{cu1} with those established in \cref{noscale19,noscale18}.
The transformation from the disc coordinates $Z$ and $Z^*$,
utilized in \Sref{cu1}, to the new ones $T$ and $T^*$ is performed
\cite{linde,old,su11} via the replacement
\beq Z
=-\sqrt{n}\frac{T-1/2}{T+1/2}~~\mbox{with}~~\re(T)>0.\label{TZ}\eeq
The last restriction -- from which the name of the $T-T^*$
coordinates -- is compatible with \Eref{logbound} for $K=\Km$.

Inserting \Eref{TZ} into \eqs{kpm}{w0pm}, $\Km$ and
$W_{0\boldsymbol-}$ may be expressed in terms of $T$ and $T^*$ as
follows
\beqs\bea \label{kmT} K_{\boldsymbol-}&=&-n\ln\frac{T+
T^*}{(T+1/2)(T^*+1/2)}\\
\mbox{and}~~W_{0\boldsymbol-}^{\pm}&=&(2T)^{\nmp}(T+1/2)^{-n},\label{W0T}
\eea\eeqs
where we fix $k=0$ in \Eref{kpm}, define the exponents
\beq n_\pm=\frac12\lf n\pm\sqrt{3n}\rg\label{npmT}\eeq
and take into account the identity
\beq
\arth\frac{Z}{\sqrt{n}}=\frac12\ln\frac{\sqrt{n}+Z}{\sqrt{n}-Z}.\label{idt}\eeq
Performing a \Kaa\ transformation as in \Eref{Kst} with
\beq \Lambda_K=-n\ln\lf T+1/2\rg\eeq
we can show that the model described by \eqs{kmT}{W0T} is
equivalent to a model relied on the following ingredients
\beq \label{WKmt} \wtilde K_{\boldsymbol-}=-n\ln\lf {T+\bar
T}\rg~~\mbox{and}~~\wtilde W_{\boldsymbol-}^{\pm}=m
(2T)^{\nmp}.\eeq
We reveal the celebrated $K$ and $W$ analyzed in \cref{noscale18,
noscale19}. Contrary to the solutions proposed in \eqs{W0}{upm},
the presence of the exponents in \Eref{WKmt} may require some
special attention from the point of view of holomorphicity
\cite{noscale18, noscale19}. Considering, though, $\wtilde
W_{\boldsymbol-}^{\pm}$ as an effective $W$, valid close to the
non-zero vacuum of the theory, any value of $n_\pm$ is, in
principle, acceptable.

Trying to achieve locally stable dS/AdS vacua with stabilized $T$
we concentrate on the following $K$
\beqs\beq \label{Kmt} \wtilde
K_{\boldsymbol-}=-n\ln\wtilde\Omega,\eeq
where the argument of $\ln$ is introduced as
\beq \label{wom} \wtilde\Omega= T+T^*+k^2\tv^4/n~~\mbox{with}~~
\tv=T+T^*-\sqrt{2}\vrm.\eeq\eeqs
As regards $W$, this can be generated by interconnecting the two
parts in \Eref{WKmt}. Namely, we define
\beq\wtilde W_{\Lambda\boldsymbol-}= \cp \wtilde
W_{0\boldsymbol-}^{+}-\cm \wtilde W_{0\boldsymbol-}^{-}
=m(2T)^{\nnp}\Ctm,\label{Wlt}\eeq
where the last short expression is achieved thanks to the new $C$
symbols defined as
\beq \Ctpm:=\cp\pm\cm (2T)^{-\sqrt{3n}}.\label{ctpm}\eeq
The resulting SUGRA potential $\wtilde V_{\boldsymbol-}$, obtained
after replacing \eqs{Kmt}{Wlt} into \Eref{Vsugra}, is found to be
\beq\begin{aligned} \wtilde V_{\boldsymbol-}&=
(m/2)^2\tom^{-n}|2T|^{2n_+}
\Big(-12|\Ctm|^2+n\tom^2\cdot\\&\left|\lf\sqrt{3}\Ctp+\sqrt{n}\Ctm\rg/T-2\sqrt{n}(1+4k^2\tv^3/n)\Ctm\tom^{-1}\right|^2\\
&\cdot\lf n\lf1+4k^2\tv^3/n\rg^2-12k^2\tv^2\tom\rg^{-1}\Big).
\end{aligned}\label{Vtm}\eeq
For the directions in \eqs{vev1}{vev2} -- with $Z$ replaced by $T$
-- we obtain dS/AdS vacua since $\veva{\wtilde
V_{\boldsymbol-}}=\vcc$ given in \Eref{Vlf}. In addition, the
conditions in \Eref{ext} for $V=\wtilde V_{\boldsymbol-}$ are
satisfied after imposing $n>3$. This is, because the sgoldstino
components ($t$ and $\bar t$) -- appearing by the decomposition of
$T$ as in \Eref{Zzzi} -- acquire masses squared
\beqs\bea \label{mt}\what
m_{t}^2&=&288\sqrt{2}k^2n^{-1}\vrm^{3}\vev{{\Ctp}/{\Ctm}}^2\mgrt^2 ;\\
\what m_{\bar
t}^2&=&4\lf1-(3/n)\vev{\Ctp/\Ctm}^2\rg\mgrt^2.\label{mbt}\eea\eeqs
Note that the expression for $\what m_{\bar t}$ coincides with
that for $\what m_{\bar z}$ in \Eref{mbzpm} for $K=\Km$ if we
replace $\Ctpm$ with $C_{u\boldsymbol-}^{\pm}$. As in that case,
to ensure $\what m_{\bar t}^2>0$ we have to impose the
aforementioned lower bound on $n$. Otherwise, an extra term of the
form $\tilde k(T-T^*)^4$ \cite{noscale18, noscale19} added in
\Eref{wom} may facilitate the stabilization for lower $n$ values.
The expressions above contain the \Gr\ mass
\beq \label{mgrt}
\mgrt=m(\sqrt{2}\vrm)^{\sqrt{3n}/2}\vev{\Ctm}\eeq
which can be determined after inserting \eqs{Kmt}{Wlt} into
\Eref{mgr}.  Upon substitution of the the mass spectrum above into
\Eref{strace} we find
\beq{\sf
STr}M_{T\boldsymbol-}^2=(12/n)\mgrt^2\vev{\Ctp/\Ctm}^2\lf24\sqrt{2}k^2\vrm^3/n-1\rg,\label{trt}\eeq
consistently with the expression of \Eref{strace1} given that
$\rcal$ from \Eref{rcal} is \beq \vev{\wtilde {\cal
R}_{T\boldsymbol-}}=\lf24\sqrt{2}k^2\vrm^3/n-1\rg/\vrm^2.\label{calrt}\eeq

Adopting the superpotential in \Eref{w0} for the visible-sector
fields $\phc_a$ and employing for simplicity $\cm\simeq0$ we below
find the resulting SSB parameters. To this end, we identify $\khi$
in \eqs{K1}{K2} with $\wtilde K_{\boldsymbol-}$ in \Eref{Kmt} and
so we obtain the corresponding $\wtilde K_{1\rm HO}$ and $\wtilde
K_{2\rm HO}$. On the other hand, $K_{3\rm HO}$ in \Eref{K3} may be
replaced with the following
\beq \wtilde K_{3\rm
HO}=-n\ln\left(\tom-\mbox{$\sum_\al$}|\phc_\al|^2/n\right).\label{tK2}
\eeq
For low $\phc_\al$ values, the $K_{\rm HO}$'s above reduce to that
shown in \Eref{Kho}, with $\wkhi$ being identified as
\beq \label{wkt} \wkhi=\begin{cases} 1&\mbox{for}~~K_{\rm
HO}=\wtilde
K_{1\rm HO}, \wtilde K_{2\rm HO};\\
\tom^{-1}&\mbox{for}~~K_{\rm HO}=\wtilde K_{3\rm
HO}\,.\end{cases}\eeq
Using the standard formalism \cite{susyr}, we extract the
following SSB masses squared
\beqs\beq \label{mti} m_\al^2=\begin{cases} \mgrt^2&\mbox{for}~~K_{\rm HO}=\wtilde K_{1\rm HO}, \wtilde K_{2\rm HO};\\
(1-3/n)\mgrt^2&\mbox{for}~~K_{\rm HO}=\wtilde K_{3\rm
HO},\end{cases}\eeq
trilinear coupling constant
\beq \label{Amt} \frac{A}{\mgrt}=
\begin{cases}-\sqrt{3n}&\mbox{for}~~K_{\rm HO}=\wtilde K_{1\rm HO}\\
&\mbox{and}~~\wtilde K_{2\rm HO};\\
\sqrt{3}\lf{3}/{\sqrt{n}}-\sqrt{n}\rg&\mbox{for}~~K_{\rm HO}=\wtilde K_{3\rm HO},\\
\end{cases}\eeq
and bilinear coupling constant
\beq \label{Bmt} \frac{B}{\mgrt}=\begin{cases}-\lf1+\sqrt{3n}\rg&\mbox{for}~~K_{\rm HO}=\wtilde K_{1\rm HO}\\
&\mbox{and}~~\wtilde K_{2\rm HO};\\
\lf\sqrt{3/n}(2-n)-1\rg&\mbox{for}~~K_{\rm HO}=\wtilde K_{3\rm
HO}.
\end{cases}\eeq\eeqs
%\vev{2T}^{\frac{3-n}{2}}
%
To reach the results above we take into account the auxiliary
expressions
\beq\begin{aligned} \vev{F^T}&=2\sqrt{3}\mgrt/\sqrt{n},
~\vev{\partial_T\kh}=-n/\sqrt{2}\vrm
\\ \vev{\partial_T\ln\wkhi^3}&={3}\vev{\partial_T\ln\wkhi^2}/2=-3/\sqrt{2}\vrm
\label{aux2ns}\end{aligned}\eeq
and define the rescaled parameters
\bea\nonumber (\what h, \what\mu) = \lf\sqrt{2}\vrm\rg^{-n/2}
(h,\mu)\eea for $K_{\rm HO}=\wtilde K_{1\rm HO}$ and $\wtilde
K_{2\rm HO}$. For $K_{\rm HO}=\wtilde K_{3\rm HO}$ we have
\bea\nonumber \what h= \lf\sqrt{2}\vrm\rg^{(3-n)/2}h~~\mbox{and}~~
\what\mu = \lf\sqrt{2}\vrm\rg^{(2-n)/2}\mu\,.\eea

For $K_{\rm HO}=\wtilde K_{3\rm HO}$ and $n=3$ we recover the
standard no-scale SSB terms as regards $m_\al$ and $A$
\cite{old,nsreview} but not for $B$ -- cf. \cref{noscaleinfl}. The
reason is that here $W$ in \Eref{Wlt} is not constant as in the
original no-scale models and this fact modifies the resulting
$\vev{F^T}$ which includes derivation of $W$ w.r.t $T$. Comparing
the above results with those in \eqs{ssbpm1}{ssbpm2} we remark
that the expressions for $m_\al$ are exactly the same.

Extensions of the present model including more than one goldstini
and also matter fields are extensively investigated in
\cref{noscale18, noscale19}.

%\newpage

\def\ijmp#1#2#3{{\sl Int. Jour. Mod. Phys.}
{\bf #1},~#3~(#2)}
\def\plb#1#2#3{{\sl Phys. Lett. B }{\bf #1}, #3 (#2)}
\def\prl#1#2#3{{\sl Phys. Rev. Lett.}
{\bf #1},~#3~(#2)}
\def\rmpp#1#2#3{{Rev. Mod. Phys.}
{\bf #1},~#3~(#2)}
\def\prep#1#2#3{{\sl Phys. Rep. }{\bf #1}, #3 (#2)}
\def\prd#1#2#3{{\sl Phys. Rev. D }{\bf #1}, #3 (#2)}
\def\npb#1#2#3{{\sl Nucl. Phys. }{\bf B#1}, #3 (#2)}
\def\npps#1#2#3{{Nucl. Phys. B (Proc. Sup.)}
{\bf #1},~#3~(#2)}
\def\mpl#1#2#3{{Mod. Phys. Lett.}
{\bf #1},~#3~(#2)}
\def\jetp#1#2#3{{JETP Lett. }{\bf #1}, #3 (#2)}
\def\app#1#2#3{{Acta Phys. Polon.}
{\bf #1},~#3~(#2)}
\def\ptp#1#2#3{{Prog. Theor. Phys.}
{\bf #1},~#3~(#2)}
\def\n#1#2#3{{Nature }{\bf #1},~#3~(#2)}
\def\apj#1#2#3{{Astrophys. J.}
{\bf #1},~#3~(#2)}
\def\mnras#1#2#3{{MNRAS }{\bf #1},~#3~(#2)}
\def\grg#1#2#3{{Gen. Rel. Grav.}
{\bf #1},~#3~(#2)}
\def\s#1#2#3{{Science }{\bf #1},~#3~(#2)}
\def\ibid#1#2#3{{\it ibid. }{\bf #1},~#3~(#2)}
\def\cpc#1#2#3{{Comput. Phys. Commun.}
{\bf #1},~#3~(#2)}
\def\astp#1#2#3{{Astropart. Phys.}
{\bf #1},~#3~(#2)}
\def\epjc#1#2#3{{Eur. Phys. J. C}
{\bf #1},~#3~(#2)}
\def\jhep#1#2#3{{\sl J. High Energy Phys.}
{\bf #1}, #3 (#2)}
\newcommand\jcap[3]{{\sl J.\ Cosmol.\ Astropart.\ Phys.\ }{\bf #1}, #3 (#2)}
\newcommand\njp[3]{{\sl New.\ J.\ Phys.\ }{\bf #1}, #3 (#2)}
\def\prdn#1#2#3#4{{\sl Phys. Rev. D }{\bf #1}, no. #4, #3 (#2)}
\def\jcapn#1#2#3#4{{\sl J. Cosmol. Astropart.
Phys. }{\bf #1}, no. #4, #3 (#2)}
\def\epjcn#1#2#3#4{{\sl Eur. Phys. J. C }{\bf #1}, no. #4, #3 (#2)}

\end{document}